\global\def\draftcontrol{0}

   \def\versionno{ kpt-(in)stability}

\catcode`\@=11

\expandafter\ifx\csname draftcontrol\endcsname\relax\global\def\draftcontrol{0}
\fi

{\count255=\time\divide\count255 by 60
\xdef\hourmin{\number\count255}
\multiply\count255 by-60\advance\count255 by\time
\xdef\hourmin{\hourmin:\ifnum\count255<10 0\fi\the\count255}}
\def\draftdate{\number\month/\number\day/\number\year\ \ \ \hourmin }

\newcommand\makepapertitle{\par
  \begingroup
    \renewcommand\thefootnote{\@fnsymbol\c@footnote}%
    \def\@makefnmark{\rlap{\@textsuperscript{\normalfont\@thefnmark}}}%
    \long\def\@makefntext##1{\parindent 1em\noindent
            \hb@xt@1.8em{%
                \hss\@textsuperscript{\normalfont\@thefnmark}}##1}%
     \newpage
     \global\@topnum\z@   
     \@makepapertitle
     \thispagestyle{empty}\@thanks
  \endgroup
  \setcounter{footnote}{0}%
  \global\let\thanks\relax
  \global\let\makepapertitle\relax
  \global\let\@makepapertitle\relax
  \global\let\@thanks\@empty
  \global\let\@author\@empty
  \global\let\@date\@empty
  \global\let\@title\@empty
  \global\let\title\relax
  \global\let\author\relax
  \global\let\date\relax
  \global\let\and\relax
  \def\version{\let\version\@version\@gobble}
}
\def\@makepapertitle{%
  \newpage
   \ifnum\draftcontrol=1 {}
   \version\versionno
   \vskip 3em%
   \else
   \hfill\hbox to 3cm {\parbox{4cm}{\@pubnum}\hss}%
   \vskip 3em%
   \fi
   \begin{center}%
   \let \footnote \thanks
     {\LARGE {\@title}}%
     \vskip 1.5em%
     {\normalsize
       \lineskip .5em%
       \begin{tabular}[t]{c}%
         \@author
       \end{tabular}\par}%
     \vskip 1.5em%
     {\@bstract}%
     \end{center}%
     \vskip 1.5em
     \@date%
   \par
}

\gdef\@pubnum{}
\def\pubnum#1{%
  \gdef\@pubnum{#1}}

\gdef\@bstract{}
\def\Abstract#1{%
  \gdef\@bstract{%
   \parbox{\textwidth-0pc}{%
   \centerline{\bf Abstract}\penalty1000%
\kern.2cm%
\noindent
\renewcommand\baselinestretch{1.0}%
{#1}}}
}

\def\ps@paper{\let\@mkboth\@gobbletwo%
     \ifnum\draftcontrol=1
    \def\@oddfoot{\hbox to \textwidth{\tiny \versionno \hfil\tiny\draftdate}%
    \hskip -\textwidth \hbox to \textwidth{\hfil\rm\thepage\hfil}}%
     \else\def\@oddfoot{\hbox to \textwidth{\hfil\rm\thepage\hfil}}
     \fi
     \let\@evenfoot\@oddfoot
}

\def\body{\clearpage
          \pagestyle{paper}
    }

\def\@version#1{\ifnum\draftcontrol=1
\typeout{}\typeout{#1}\typeout{}
\vskip3mm\centerline{\hbox{\fbox{\normalsize{\tt DRAFT -- #1 -- }
                   {\draftdate}}}}\vskip3mm
\fi}
\let\version\@version
\long\def\eqlabel#1{\ifnum\draftcontrol=1
                    \tag@false  
                    \tag*{(\theequation) \hbox to -0.2cm{\hspace{0cm}\small{#1}\hss}}
                    \refstepcounter{equation}
                    \edef\@currentlabel{\theequation}
                    \ltx@label{#1}          
                    \else
                    \label{#1}
                    \fi
                    }
\let\st@bibitem\@bibitem
\let\st@lbibitem\@lbibitem
\ifnum\draftcontrol=1
  \def\@bibitem#1{%
    \st@bibitem{#1}\a@@label{#1}\ignorespaces}
  \def\@lbibitem[#1]#2{%
    \st@lbibitem[#1]{#2}\a@@label{#2}\ignorespaces}
  \def\a@@label#1{%
    \gdef\a@lab{\smash{\normalfont\small#1}}
    \ifvmode
      \if@inlabel
        \global\setbox\@labels\hbox{%
          \llap{\a@lab\let\a@lab\relax
                \kern\@totalleftmargin\kern\marginparsep}%
          \box\@labels}%
      \fi
    \fi}
\fi

\documentclass[12pt,letterpaper]{article}

\usepackage{amsmath,amssymb,array,calc,epsfig,rotating,bm}
\usepackage[sort]{cite}
\usepackage{graphicx}
\usepackage{psfrag,verbatim}
\usepackage{xcolor}
\usepackage{hyperref}

\ifnum\draftcontrol=1
\fi

\renewcommand\baselinestretch{1.25}
\renewcommand\section{\@startsection {section}{1}{\z@}%
                                   {-3.5ex \@plus -1ex \@minus -.2ex}%
                                   {2.3ex \@plus.2ex}%
                                   {\normalfont\large\bfseries}}
\renewcommand\subsection{\@startsection{subsection}{2}{\z@}%
                                   {-3.25ex\@plus -1ex \@minus -.2ex}%
                                   {1.5ex \@plus .2ex}%
                                   {\normalfont\normalsize\bfseries}}
\renewcommand\subsubsection{\@startsection{subsubsection}{3}{\z@}%
                                   {-3.25ex\@plus -1ex \@minus -.2ex}%
                                   {1.5ex \@plus .2ex}%
                                   {\normalfont\normalsize\it}}
\renewcommand\paragraph{\@startsection{paragraph}{4}{\z@}%
                                   {-3.25ex\@plus -1ex \@minus -.2ex}%
                                   {1.5ex \@plus .2ex}%
                                   {\normalfont\normalsize\bf}}

\numberwithin{equation}{section}

\def\revise#1       {\raisebox{-0em}{\rule{3pt}{1em}}%
                     \marginpar{\raisebox{.5em}{\vrule width3pt\
                     \vrule width0pt height 0pt depth0.5em
                     \hbox to 0cm{\hspace{0cm}{%
                     \parbox[t]{4em}{\raggedright\footnotesize{#1}}}\hss}}}}

\newcommand\nxt[1]  {\\\fnxt#1}
\newcommand{\ie}{{\it i.e.,}\ }

\def\cala         {{\cal A}}

\def\calb         {{\cal B}}

\def\cald         {{\cal D}}

\def\calf         {{\cal F}}
\def\calg         {{\cal G}}

\def\calk         {{\cal K}}
\def\call         {{\cal L}}
\def\calm         {{\cal M}}
\def\caln         {{\cal N}}
\def\calo         {{\cal O}}

\def\complex      {{\mathbb C}}

\def\projective   {{\mathbb P}}

\def\reals        {{\mathbb R}}

\def\del          {\partial}

\def\Re           {{\rm Re\hskip0.1em}}
\def\Im           {{\rm Im\hskip0.1em}}

\def\sqr#1#2{{\vcenter{\vbox{\hrule height.#2pt
 \hbox{\vrule width.#2pt height#1pt \kern#1pt
 \vrule width.#2pt}\hrule height.#2pt}}}}

\newcommand{\kk}{\mathfrak{q}}
\newcommand{\ww}{\mathfrak{w}}

\def\aa1{\phi}
\def\cc1{\psi}

\def\k{\kappa}

\def\k{\kappa}

\catcode`\@=12

\begin{document}


\title{\bf Near-extremal membranes in M-theory}

\date{November 3, 2025}

\author{
Alex Buchel$ ^1$ and Ruben Monten$ ^2$\\[0.4cm]
\it $ ^1$Department of Physics and Astronomy\\ 
\it University of Western Ontario\\
\it London, Ontario N6A 5B7, Canada\\
\it $ ^2$ Theoretical Physics Department, CERN\\
\it 1211 Geneva 23, Switzerland
}

\Abstract{We consider near-extremal membranes embedded in M-theory,
consistently truncated to gauged $\mathcal{N} = 2$ supergravity in four dimensions on
the coset space $M^{1,1,0}$. These are holographically dual to $2 + 1$ dimensional
superconformal gauge theory with $U(1)_R \times U(1)_B$
global symmetry. Turning on the chemical potential to either the
$R$-symmetry or the baryonic symmetry gives access to the quantum critical
regime of the boundary gauge theory.
We study perturbative stability of the extremal limit,
and demonstrate that membranes with topological (baryonic) charge are free from 
all known instabilities. $R$-charged membranes
are free from the superconducting instabilities, but have unstable charge transport
and instabilities associated with the condensation of the axions. 
}

\makepapertitle
\bigskip
\hfill CERN-TH-2025-224

\body

\version\versionno
\tableofcontents

\section{Introduction and summary}\label{intro}
Holographic correspondence \cite{Maldacena:1997re,Aharony:1999ti} provides
an interesting realization of quantum criticality: seemingly violating the
third law of thermodynamics, a strongly coupled phase of
matter in the limit of vanishing temperature $T\to 0$ has finite entropy density,
while lacking the supersymmetry. In the gravitational dual, such phases
are described by charged black branes in string theory/M-theory with non-supersymmetric
extended extremal horizons. The best explored holographic example
is that of the strongly coupled $\caln=4$ supersymmetric Yang-Mills (SYM) plasma
in four spacetime dimensions.
Here,  the equilibrium states of the gauge theory plasma,
with the same chemical potential $\mu$ for all $U(1)$ factors of the maximal
Abelian subgroup $U(1)^3\subset SU(4)$ $R$-symmetry, reach the quantum critical
regime as $\frac{T}{\mu}\to 0$. In the gravitational dual, such states are
represented by a Reissner--Nordström (RN) black brane in asymptotically
$AdS_5$ spacetime. 

A possible resolution of the extremal entropy paradox
is the modification of the $T\to 0$ limit by the quantum gravity corrections
\cite{Turiaci:2023wrh}. As such corrections only become important at exponentially
suppressed temperatures, it is natural to ask if a more mundane resolution exists.
Indeed, one can only sensibly talk about quantum gravity corrections in
a consistent theory of quantum gravity --- the string theory. Embedding
extremal horizons in consistent string  theory backgrounds
(in the classical supergravity approximation) typically involves plethora of
additional charged and neutral fields. It is then possible
that the extremal limit is never reached due to various perturbative
and non-perturbative instabilities, triggered by these spectator fields. 
Generic examples are the holographic superconducting instabilities
\cite{Hartnoll:2008vx}, associated with the condensation of the
operators (in the gravitational dual bulk scalars) charged under the global $U(1)$ symmetry that supports the extremal limit.%
\footnote{%
We will use the term “superconducting” instability exclusively when \emph{charged} bulk fields condense. 
We will also encounter the condensation of the (neutral) axions at, what we refer to as, “threshold” instabilities.
}
In the case of $\caln=4$ SYM such an
operator is a chiral primary gaugino bilinear \cite{Gubser:2009qm,Buchel:2025ves}.

More subtle examples are extremal horizons supported by gauge fields
realizing ``topological'' global symmetries of the boundary gauge theory.
Such holographic models arise from compactifications of string theory/M-theory
on $AdS_{p+2}\times Y$ manifolds with nonzero $p$th Betti number $b_p$, leading to $U(1)^{b_p}$ “baryonic” global symmetry.
Non-supersymmetric extremal quantum states supported by the baryonic  $U(1)^{b_p}$ 
chemical potentials do not have superconducting instabilities. As an example,
consider strongly coupled $\caln=1$ $SU(N)\times SU(N)$ gauge theory in four
spacetime dimensions, the Klebanov--Witten (KW) model \cite{Klebanov:1998hh}.
The theory has $U(1)_R\times U(1)_B$ global symmetry, which supports
quantum critical states charged under either of the $U(1)$s.
The $R$-symmetry charged quantum critical states are unstable 
due to the condensation of the chiral primary
$\calo_F\equiv {\rm Tr} (W_1^2+W_2^2)$, where $W_i$ are the gauge superfields corresponding
to the two gauge group factors of $SU(N)\times SU(N)$ quiver \cite{Buchel:2024phy}. 
The gauge-invariant operators  of the KW theory charged under $U(1)_B$ have conformal
dimensions of order\footnote{The smallest such operators involve determinants of the
bifundamental matter fields of the KW quiver gauge theory. This justifies the nomenclature “baryonic symmetry”.} $N$, with the charge-to-mass ratio
too small to trigger the superconducting instability \cite{Herzog:2009gd}. 
Nonetheless, quantum critical states with a baryonic charge of the KW theory are
unstable \cite{Buchel:2025jup}:
even though such states have zero $R$-symmetry charge density, at low temperatures
$R$-charge starts
“clumping”, breaking the homogeneity of $U(1)_B$ charged
thermal equilibrium state.\footnote{This is a direct consequence of the
thermodynamic instabilities of the underlying thermal states \cite{Buchel:2005nt}. For charged plasma this was originally explained in
\cite{Gladden:2024ssb,Gladden:2025glw}.
}

In this paper we study extremal horizons in a close analog of the 
KW model --- a membrane theory of Klebanov, Pufu and
Tesileanu (KPT) \cite{Klebanov:2010tj}. The KPT model
is a holographic example of a three dimensional superconformal
gauge theory arising from compactification of M-theory
on regular seven-dimensional Sasaki--Einstein manifold
with fluxes \cite{Cassani:2012pj}. The full list of such manifolds 
is given in \cite{Friedrich:1990zg}, and we focus on a particular example.
The starting point is $\frac{SU(2)^3}{U(1)^2}$ coset, known as $Q^{1,1,1}$,
which is a $U(1)$ fibration over $\complex\projective^1\times\complex\projective^1\times\complex\projective^1$. This manifold has second Betti number $b_2=2$, and we further
consistently truncate one Betti vector of $Q^{1,1,1}$ to
arrive\footnote{Baryonic black membrane theory of KPT is
precisely such compactification.} at
$\frac{SU(3)\times SU(2)}{SU(2)\times U(1)}$ coset, known as
$M^{1,1,0}$, with a single topological $U(1)$. 
Much like the KW theory, the
holographic membrane model of M-theory on $M^{1,1,0}$ has
$U(1)_R\times U(1)_B$ global symmetry. We show that there are 3 distinct
near-extremal regimes: one supported by the $U(1)_R$ charge density,
and the other two supported by the $U(1)_B$ charge density.
The reason for the distinct baryonic near-criticality comes from the
fact that the dual gravitational backgrounds
have nontrivial support from the bulk scalar with $m^2 L^2=-2$,
corresponding to an operator of conformal dimension $\Delta=(2,1)$.
Depending on whether one uses normal or alternative quantization \cite{Klebanov:1999tb}, one obtains either of two field theory duals,
each with a near-extremal regime.

We now report the results of
our extensive analysis of the perturbative stability of the model:
\begin{itemize}
\item $U(1)_R$ quantum criticality:
\nxt While there is a single $R$-charged operator
in the theory of conformal dimension $\Delta=5$, its $R$-charge
is too small to cause its condensation at the extremality.
\nxt We study baryonic charge transport and demonstrate that the
$U(1)_B$ diffusion coefficient $D_B$ becomes negative below some critical
temperature $T_{crit}$, relative to the $R$-charge chemical potential $\mu_R$
of the near-critical thermal equilibrium states,
\begin{equation}
\begin{cases}
D_B>0\,,\qquad {\frac{T}{\mu_R}> \frac{T}{\mu_R}\bigg|_{crit}}\,,\\
D_B<0\,,\qquad {\frac{T}{\mu_R}< \frac{T}{\mu_R}\bigg|_{crit}}\,.
\end{cases}
\eqlabel{dinb}
\end{equation}
This triggers an instability of a diffusive mode in the hydrodynamic
sound channel with a dispersion relation\footnote{We use notations
$\ww\equiv\frac{w}{2\pi T}$ and $\kk\equiv \frac{|\vec k|}{2\pi T}$ where $e^{-i w t + i \vec k \cdot \vec x}$ is the profile of the hydrodynamic perturbation.}
\begin{equation}
\ww=-i D_B\kk^2+\calo(\kk^2)\,,
\eqlabel{diffinb}
\end{equation} 
resulting in the spatial baryonic charge clumping.
The precise value of the critical temperature depends on what
quantization condition is used for gravitational dual
bulk pseudoscalar coupling the electric components of the $U(1)_B$
gauge field with the magnetic components of the $U(1)_R$
gauge field. The critical temperature is larger if the pseudoscalar
is quantized so that the dual operator has a conformal dimension
$\Delta=1$.
\nxt The pseudoscalars mentioned above are neutral under the
$U(1)_R$ symmetry. However, we show that there is a critical
temperature at which their homogeneous and isotropic
fluctuations become normalizable --- this signals an onset
of the instability, potentially leading to new low-temperature
phases of the model\footnote{The detailed
exploration of these phases will be reported elsewhere.}.
Once again, the critical temperature here depends on the
pseudoscalar quantization. For both the normal and
the alternative quantizations it is lower than the corresponding
critical temperature for the $U(1)_B$ charge clumping instability \eqref{dinb}, 
which sets in first when the temperature is lowered.
\item $U(1)_B$ quantum criticality (the KPT model):
\nxt Since there are no fields of dimension $\Delta \sim \calo(1)$ charged
under $U(1)_B$ symmetry,  there can not be perturbative
superconducting instabilities of the model.
\nxt We study $U(1)_R$ charge transport of the model ---
in all cases, and for all values of $\frac{T}{\mu_B}$, the diffusion
coefficient $D_R>0$. In the extremal $T\to 0$
limit it either remains constant, $2\pi T D_R\propto +\frac{T}{\mu_B}$,
or vanishes, $2\pi T D_R\propto +\left(\frac{T}{\mu_B}\right)^2$,
depending on what quantization is used for scalars supporting
the baryonic black membrane background, as well as what quantization is
used for pseudoscalars (see section \ref{fluckpt} for further details).
\nxt There are no threshold instabilities associated with the
pseudoscalar condensation in the model (see section \ref{axionrn}
for further details).
\end{itemize}
Our main conclusion is that the KPT model \cite{Klebanov:2010tj}
is free from perturbative
instabilities in the (exotic --- \ie with finite entropy
density) quantum critical regime. Since it was already
checked in \cite{Klebanov:2010tj} that the model is
free\footnote{We would like thank Igor Klebanov
for emphasizing this point.} from the non-perturbative “Fermi seasickness”
instability \cite{Hartnoll:2009ns}, it appears to
be the first example of the classically stable
non-supersymmetric extremal horizon in string theory/M-theory.

The rest of the paper is organized as follows.
In section \ref{effact} we follow \cite{Cassani:2012pj} (CKV)
and review the consistent truncation of M-theory on the
$Q^{1,1,1}$ coset. We find it convenient to follow
\cite{Louis:2002ny} and further dualize the massive
2-form $B$ (in the expansion of the 11-dimensional
3-form gauge potential $A_3$ \eqref{a3})
to a massive vector $A^H$. Section \ref{bbm} 
deals with the baryonic black membranes. In section \ref{kpt} we start 
with  the CKV
effective action, see \eqref{kpt1}, and
reproduce the baryonic black membrane
solution of \cite{Klebanov:2010tj}. 
Note that the solution is supported by the non-trivial profiles
of the two bulk scalars $v_1$ and $v_2$. The boundary gauge
theory operator dual to $\ln\frac{v_1}{v_2}$
can be identified either with an operator $\calo_2$ (of the conformal
dimension $\Delta=2$) or $\calo_1$ (of the conformal dimension
$\Delta=1$). Only the former quantization was used in \cite{Klebanov:2010tj}.
We construct black membrane solutions in both
quantizations in section \ref{backpt}. As expected, the quantum criticality
(precise $T=0$ geometry) is identical for either of the quantizations,
but the near-extremal limits, $\frac{T}{\mu_B}>0$, differ. 
In section \ref{fluckpt} we identify a decoupled
set of fluctuations associated with the $R$-charge transport:
besides the excitation of the electric components of the $A^0$
gauge field,
dual to a conserved $R$-symmetry current of the boundary
gauge theory \cite{Gladden:2024ssb}, one must include
the fluctuations of the
massive gauge field $A^H$, the magnetic components
of $A^1-A^2$, dual to a conserved baryonic symmetry
current, and the pseudoscalars (the axions) $b_1$ and $b_2$
arising from the expansion of the 11D supergravity
3-form $A_3$, see \eqref{a3}.
The latter naturally combine with the scalars $v_i$ into
complex scalars $t_i=v_i+i b_i$, see \eqref{actionckv}. 
Like for the $v_i$, one can impose
different quantizations for the axions, leading
to the identification of the linearized fluctuation $(b_1-b_2)$ 
with either $\delta\calo^b_2$ or $\delta\calo^b_1$ operators.
We compute the $R$-charge dimensionless diffusion coefficient
$\cald_R\equiv 2\pi T D_R$
for all four possible quantizations $\{\ln\frac{v_1}{v_2}\,, b_1-b_2\}\
\Longleftrightarrow\ \{\calo_2\,,\delta\calo^b_2\}$, or 
$\{\calo_2\,,\delta\calo^b_1\}$, $\{\calo_1\,,\delta\calo^b_2\}$,
$\{\calo_1\,,\delta\calo^b_1\}$, see fig.~\ref{figure3}.
In section \ref{axionkpt} we study the decoupled
set involving the homogeneous fluctuations of $(b_1-b_2)$
and demonstrate the absence of the threshold instabilities. 
We move to discussing $R$-charge supported quantum criticality
of our model in section \ref{rnbm}. In section \ref{rnbackground}
we derive Reissner–Nordstr\"om black membrane solution
from the CKV effective action:
a peculiar feature is the necessity to turn on both the
electric component of $A^0$ gauge field and the magnetic component
of the $A^1=A^2$ gauge fields \cite{Gauntlett:2009zw}
(see \eqref{f0f1}) in order to decouple
the massive vector $A^H$ and the axions $b_i$.  
Section \ref{rnfluc} is a detailed analysis of the
baryonic charge transport in this R-charged background. The structure of the decoupled
set of linearized fluctuations closely resembles
the discussion of that in section \ref{fluckpt}
with the roles of the two massless bulk gauge fields reversed
$A^0\Leftrightarrow (A^1-A^2)$. Once again, consistency of the
truncation requires the excitation of the massive gauge field
$A^H$ and the axions $b_i$. Since the scalars $v_i$ are trivial in
the RN background,  $v_i\equiv 1$ \eqref{rn}, 
there only two different cases for the diffusion coefficient
of the baryonic charge transport, depending on what quantization
we choose for the fluctuations $(b_1-b_2)$: $\delta\calo^b_2$
or $\delta\calo^b_1$. As we indicated earlier, the baryonic
charge transport in the RN membrane background is unstable
at low temperatures, see \eqref{dinb} and fig.~\ref{figure4}.
Additional instabilities in the model are associated
with the homogeneous and isotropic
condensation of the neutral pseudoscalar $(b_1-b_2)$,
although these instabilities occur at  lower temperatures than
the corresponding critical temperatures for the baryonic charge clumping, 
see section \ref{axionrn}. Finally, in section
\ref{rnsuper} we consider potential
holographic superconductor \cite{Hartnoll:2008vx} instabilities
in the model: CKV effective action includes a complex bulk scalar
$\chi\equiv \frac{1}{\sqrt{3}}(\xi^0+i\tilde\xi^0)$ of conformal
dimension $\Delta_\chi=5$
with\footnote{Our conversion for the $R$-charge normalization
differs by a factor of 2 from the one used in \cite{Cassani:2012pj}.}
$R(\chi)=4$. Since this is the only $R$-charged field, its
linearized fluctuations decouple from the other fields of the model. 
We explicitly verified that $\chi$ does not condense. This agrees with 
the comprehensive probe analysis in \cite{Denef:2009tp} once we
appropriately match the conventions.

In this paper we analyzed perturbative stability of the
extremal horizons of  M-theory compactified on $M^{1,1,0}$.
The stability analysis is specific to the model, and
thus it is interesting to extend the quest for
classically stable extremal horizons
to other examples of Sasaki-Einstein manifolds
\cite{Friedrich:1990zg}. Of course, one must
keep an open mind for additional instabilities
that eluded the current analysis.

\section{Effective action}\label{effact}

We follow notations of \cite{Cassani:2012pj} and review
the consistent truncations of 11D supergravity on seven-dimensional
Sasaki--Einstein coset $M^{1,1,0}$ with fluxes. The resulting $\caln=2$
gauged supergravity in four dimensions embeds the
holographic duality of the KPT membrane model \cite{Klebanov:2010tj}
with $U(1)_R\times U(1)_B$ global symmetry. 

The starting point is 11D supergravity
\begin{equation}
S_{11}=\frac{1}{2 \kappa_{11}^2}\int_{\calm_{11}}
\left(R\star 1 -\frac 12 G_4 \wedge\star G_4-\frac 16 A_3 \wedge G_4 \wedge G_4\right)\,.
\eqlabel{11d}
\end{equation}
We consider consistent truncation of \eqref{11d} on the coset $Q^{1,1,1}$:
\begin{itemize}

\item The 11D metric is
\begin{equation}
ds^2=e^{2V}\calk^{-1}\ ds_4^2 + e^{-V} \sum_{i=1}^3 \frac{v_i}{8}
\biggl(\ (d\theta_i)^2+\sin^2\theta_i (d\phi_i)^2\ \biggr)+e^{2 V }\biggl(
\theta+ A^0\biggr)^2 \ ,
\eqlabel{met11}
\end{equation}
where $ds_4^2$ is the 4D metric, $A^0$ is a 1-form on $\calm_4$, and where
\begin{equation}
\begin{split}
&\calk \equiv v_1 v_2 v_3 \ ,
\qquad \theta \equiv d\psi+\frac 14 \sum_{i} \cos\theta_i d\phi_i \ . 
\end{split}
\end{equation}
We will furthermore define the scalar field $\phi = \frac32 V - \frac12 \sum_i \ln v_i$ as well as the 2-forms $\omega_i = \frac18 \sin \theta_i d\phi_i \wedge d\theta_i$ so that $d\theta = \sum_i m_i \omega_i$ with $m_i=\{2,2,2\}$. 
\item The fluxes are
\begin{equation}
A_3=C_3 +B\wedge (\theta+A^0)-A^i\wedge \omega_i+b_i \omega_i\wedge (\theta+A^0)\,,
\eqlabel{a3}
\end{equation}
with $C_3$, $B$, $A^i$ and $b_i$ being correspondingly
the $3-$, $2-$, $1-$ and the $0-$ forms on $\calm_4$. 
\end{itemize}

Under this ansatz, the action \eqref{11d} reduces into the following components: 
\begin{itemize}
\item The 11D Einstein--Hilbert term becomes
\begin{equation}
S_{EH}=\frac{1}{2 \kappa_{11}^2}\int_{\calm_{11}}\ R\star 1\equiv
\frac{1}{\kappa_4^2}\ \int_{\calm_4}\biggl[\frac 12 R_4\star 1+\call_{kin,geo}
-V_{geo}\star 1\biggr] \,,
\eqlabel{eh4}
\end{equation}
where $\kappa_4^{-2}=\kappa_{11}^{-2}\int_{Q^{1,1,1}} d\psi \wedge \omega_1 \wedge \omega_2 \wedge \omega_3$ and 
\begin{align}
\call_{kin,geo} &= - (\del\phi)^2 \star 1 + \frac14 \sum_i (\del \ln v_i)^2 \star 1
-\frac 14 \calk\ \calf^0
\wedge \star \calf^0\,,
\eqlabel{kingeo}
\\
V_{geo} &= e^{4\phi} v_1v_2v_3\cdot\sum_i v_i^{-2}-8
e^{2\phi}\cdot\sum_i v_i^{-1} \ ,
\eqlabel{vgeo}
\end{align}
where $\calf^0 \equiv d A^0$.
\item The kinetic flux term can be written as
\begin{equation}
	-\frac1{4\k_{11}^2} \int_{\calm_{11}} G_4 \wedge \star G_4 = \frac1{\k_4^2} \int_{\calm_4} \left( \call_{kin, flux} - V_{flux} \star 1 \right) \,.
\end{equation}
In the following, we denote the field strengths as $\calf^I \equiv d A^I$ and the generalized field strengths as $F^I = \calf^I - m^I B$, with $I=\{0,1,2,3\}$ and $m^I=\{0,2,2,2\}$. We then get
\begin{equation}
\begin{split}
&\call_{kin,flux}=-\frac 14 \sum_i v_i^{-2}\ (\del b_i)^2\star 1-\frac 14 e^{-4\phi}
dB \wedge \star dB\\&\qquad\qquad\qquad +\frac 14 \left(\Im \caln_{IJ}
+\calk\delta^0_I\delta^0_J\right)\
F^I\wedge \star F^J+\frac 14 \Re\caln_{IJ}\ F^I\wedge F^J \,,\\
&\Re \caln_{00}=-\frac 13 \calk_{ijk}b_ib_jb_k\,,\qquad \Re \caln_{0i}=\frac12
\calk_{ijk} b_jb_k\,,\qquad \Re\caln_{ij}=-\calk_{ijk}b_k\,,\\
&\Im \caln_{00}=-\calk(1+4 g_{ij}b_ib_j)\,,\qquad \Im \caln_{0i}=4\calk g_{ij}b_j\,,\qquad \Im\caln_{ij}=-4\calk g_{ij}\,,
\end{split}
\eqlabel{flux}
\end{equation}
where $\calk_{ijk}=1$ for $i\ne j\ne k$ and $0$ otherwise, and $g_{ij}=\frac 14
v_i^{-2}\ \delta_{ij}$. The potential $V_{flux}$ is
\begin{equation}
V_{flux} = \frac{e^{4\phi}}{4 \calk}\cdot
\sum_{k} \biggl[\sum_{ij}\ \calk_{ijk}\ b_im_j v_k \
\biggr]^2 +\frac {e^{4\phi}}{4} \calk^{-1}\cdot \biggl[ e_0 + \frac 12 \sum_{i,j,k} \calk_{ijk}\ b_ib_jm_k\biggr]^2\,,
\eqlabel{vflux}
\end{equation}
where we dualized $dC_3$ as in \cite{Louis:2002ny},
\begin{equation}
\frac{e^{-4\phi}\calk}{2}\ \star\biggl(dC_3+B\wedge F^0\biggr)=
-\frac12\biggl( e_0+\frac 12 \sum_{i,j,k} b_ib_jm_k \calk_{ijk}\biggr)
\eqlabel{sdc3} \ .
\end{equation}
The constant $e_0$ will set the radius of the asymptotic $AdS_4$ spacetime. Below, we will choose $e_0 = 6 \Rightarrow L = 1/2$.
\item The topological term is
\begin{equation}
-\frac 1{12 \k_{11}^2} \int_{\calm_11} A_3 \wedge G_4 \wedge G_4 
= \frac1{\k_4^2} \int_{\calm_4} \call_{top}
= -\frac{e_0}{2\k_4^2} \int_{\calm_4} dB\wedge A^0 \ .
\eqlabel{top}
\end{equation}
\end{itemize}
Combining the gravitational and flux contributions, we reproduce 
exactly the effective action of \cite{Cassani:2012pj}:
\begin{equation}
\begin{split}
&S_{CKV}=\frac {1}{\kappa_4^2}\int_{\calm_4}
\biggl[
\frac 12 R_4\star 1-\biggl\{\ (\del\phi)^2+g_{ij} \del t^i\del \bar{t}^j
\ \biggr\}
\star 1 
-\frac 14 e^{-4\phi} dB\wedge\star dB\\
&\qquad +\frac 14\Im
\caln_{IJ} F^I\wedge\star F^J+\frac 14\Re\caln_{IJ} F^I\wedge F^J-\frac 12 e_0\ dB\wedge A^0
-V_{CKV}\star 1\biggr]\,,\\
&V_{CKV}=e^{4\phi} \calk\cdot\sum_i v_i^{-2}-8
e^{2\phi}\cdot\sum_i v_i^{-1}+\frac{e^{4\phi}}{4} \calk^{-1}\cdot\
\sum_{k} \biggl[\sum_{ij}\ \calk_{ijk}\ b_im_j v_k\
\biggr]^2
\\&\qquad +\frac {e^{4\phi}}{4} \calk^{-1}\cdot
\biggl[
e_0+\frac 12 \sum_{i,j,k} \calk_{ijk}\ b_ib_jm_k\biggr]^2\,,
\end{split}
\eqlabel{actionckv}
\end{equation}
with $t^i\equiv v_i+i b_i$.

Note that the effective action \eqref{actionckv} is invariant
under the 1-form gauge transformations (with the 0-form
gauge parameters $\alpha^0, \alpha^i$):
\begin{equation}
A^0\to A^0+d\alpha^0\,,\qquad A^i\to A^i+d\alpha^i\,.
\eqlabel{gauge}
\end{equation}

For the next step we would like to follow \cite{Louis:2002ny}
and dualize the massive 2-form $B$ to a massive vector $A_H$.%
\footnote{Here we deviate from the logic of \cite{Gauntlett:2009zw,Cassani:2012pj}, which dualized the 2-form while preserving the standard matter-coupled $\caln = 2$ Lagrangian form by first performing an electric-magnetic symplectic transformation on the vector fields. Instead we prefer to keep the gauging magnetic.}
They treat an action of the general form
\begin{equation}
\begin{split}
\call_B=-\biggl[
h\ dB\wedge \star dB+M^2\ B\wedge \star B+M_T^2\ B\wedge B+B\wedge J_2
\biggr]\,,
\end{split}
\eqlabel{baction}
\end{equation}
where in our case:
\begin{equation}
\begin{split}
&h=\frac 14 e^{-4\phi}\,,\qquad M^2 =-\frac14  \Im \caln_{IJ}\ m^I m^J\,,\qquad
M^2_T =-\frac14  \Re \caln_{IJ}\ m^I m^J\,,\\
&J_2=J_a+\star J_b\,,\\
&J_a=-\frac 12 e_0\ \calf^0 +\frac 14 \Re \caln_{IJ}\ \left(m^I\calf^J+m^J
\calf^I\right)\,,
\qquad J_b=\frac 14 \Im \caln_{IJ}\ \left(m^I\calf^J+m^J \calf^I\right)\,.
\end{split}
\eqlabel{b2}
\end{equation}
Introducing a massive vector $A^H=h\ \star dB $ we rewrite \eqref{baction}
as \cite{Louis:2002ny}
\begin{equation}
\begin{split}
-\call_B=-\call_{AH}\equiv& \frac 1h A^H\wedge\star A^H+\frac{M^2}{M^4+M_T^4}
\left(dA^H-\frac 12 J_2\right)\wedge\star \left(dA^H-\frac 12 J_2\right)
\\
&-\frac{M_T^2}{M^4+M_T^4}
\left(dA^H-\frac 12 J_2\right)\wedge \left(dA^H-\frac 12 J_2\right)\,.
\end{split}
\eqlabel{ahaction}
\end{equation}

Finally, from here on we consider the trivial consistent sub-truncation $Q^{1,1,1}\to M^{1,1,0}$ which means identifying
\begin{equation}
A^3\equiv A^1\,,\qquad v_3\equiv v_1\,,\qquad b_3\equiv b_1 \,.
\eqlabel{subtr}
\end{equation}

\section{Baryonic black membranes}\label{bbm}

\subsection{Truncation to KPT}\label{kpt}

In this section we describe the truncation of the effective action
of section \ref{effact}
to the one used in \cite{Klebanov:2010tj}.
We emphasize that this is a truncation of M-theory membranes with topological
charge for the equilibrium
thermal homogeneous solutions \underline{only}, and is inconsistent at the level of fluctuations.  
The solutions considered in \cite{Klebanov:2010tj} are homogeneous and isotropic
black membranes of 11D supergravity on
$AdS_4\times M^{1,1,0}$ with a {\it baryonic} chemical potential:
\begin{itemize}
\item The 11D metric is a warped product of $\calm_4$ and a squashed $M^{1,1,0}$
\begin{equation}
\begin{split}
ds^2=&e^{-7\chi/2}\ ds_{4}^2+4 L^2 e^{\chi}
\biggl[
\frac{e^{\eta_1}}{8}\ \sum_{i=1,3} \biggl(\
(d\theta_i)^2+\sin^2\theta_i (d\phi_i)^2\ \biggr)\\
&+\frac{e^{\eta_2}}{8}\ \biggl(\
(d\theta_2)^2+\sin^2\theta_2 (d\phi_2)^2\ \biggr)+e^{-4\eta_1-2\eta_2}\
\theta^2
\biggr]\,, 
\end{split}
\eqlabel{kptm}
\end{equation}
\begin{equation}
ds_4^2=-\calg e^{-w}\ dt^2+\frac{r^2}{L^2}\left[d(x_1)^2+d(x_2)^2\right]
+\frac{dr^2}{\calg}\,;
\eqlabel{kptm4}
\end{equation}
\item The 4-form flux $G_4$ is given by\footnote{We changed
the overall sign of $G_4$ for consistency with the action of section
\ref{effact}.}
\begin{equation}
G_4=\frac3L e^{-\frac{21}{2}\chi}\ \star_4 1 -8 Q L^3\ \frac{e^{-\frac w2-\frac 32\chi}}
{r^2}\ dt\wedge dr\wedge \biggl( e^{2\eta_1}(w_1+w_3)-2 e^{2\eta_2}w_2\biggr)\,,
\eqlabel{kptf}
\end{equation}
\end{itemize}
where the $\calm_4$ metric warp factors $\calg,w$ and the bulk scalars
$\chi,\eta_i$ are functions of the radial
coordinate $r$ only. Furthermore, $L$ is the radius of the asymptotic $AdS_4$ and 
$Q$ is the baryonic charge of the black membranes.

To compare with the notation of the previous section, we begin by comparing the metric \eqref{kptm} to \eqref{met11}, which implies $A^0 = 0$ as well as 
\begin{equation}
e_0=6 \qquad \Longleftrightarrow\qquad L=\frac 12\,,
\end{equation}
and 
\begin{align}
\chi &= -\frac8{21} \phi + \frac4{21} \ln v_1 + \frac2{21} \ln v_2
\ , \nonumber \\
\eta_1 &= -\frac27 \phi + \frac17 \ln v_1 - \frac37 \ln v_2
\ , \\
\eta_2 &= -\frac27 \phi - \frac67 \ln v_1 + \frac47 \ln v_2
\ . \nonumber
\end{align}
Matching $dA_3$ in \eqref{a3} with \eqref{kptf} furthermore requires
\begin{equation}
b_i\equiv 0 \ ,
\qquad B \equiv 0
\ ,
\eqlabel{b0}
\end{equation}
(i.e.  $A^H = 0$) as well as $\calf^1\wedge \calf^2=0$; however, this is consistent only if $J_2$ in \eqref{baction} vanishes as well. This, in turn, implies
\begin{equation}
J_2\equiv 0\qquad \Longrightarrow\qquad \frac14\Im \caln_{ij}\ m_i \calf^j\equiv 0
\qquad \Longleftrightarrow\qquad \sum_i\frac{\calf^i}{v_i^2}=0
\ .
\eqlabel{j20}
\end{equation}
As we show shortly, the last equality in \eqref{j20} is indeed satisfied
on solutions \eqref{kptm}-\eqref{kptf}
of \cite{Klebanov:2010tj}, but alas, it can not be imposed
at the level of fluctuations. 

Under these identifications, the effective action \eqref{actionckv} takes the form
\begin{equation}
\begin{split}
&\call_{CKV}\to \call_{KPT}=\frac 12 R_4 -(\del\phi)^2-\frac 12
(\del\ln v_1)^2-\frac 14 (\del\ln v_2)^2-\frac {v_2}{4}\ \calf^1_{\mu\nu}\calf^{1\mu\nu}\\
&\qquad\qquad\qquad\qquad-\frac{v_1^2}{8 v_2}\ \calf^2_{\mu\nu}\calf^{2\mu\nu}-V_{KPT}\,,
\end{split}
\eqlabel{kpt1}
\end{equation}
with
\begin{equation}
V_{KPT}=e^{4\phi} \biggl[
\frac{e_0^2}{4v_1^2 v_2}+v_1^2 v_2 \left(\frac{2}{v_1^2}+\frac{1}{v_2^2}\right)
\biggr]-8 e^{2\phi}\left(\frac{2}{v_1}+\frac{1}{v_2}\right)\,,
\eqlabel{kpt3}
\end{equation}
subject to additional constraint \eqref{j20}
\begin{equation}
\frac{2\calf^1}{v_1^2}+\frac{\calf^2}{v_2^2}=0\,.
\eqlabel{kpt2}
\end{equation}
This constraint is consistent with the 2-form equations of motion derived from \eqref{kpt1}, which read
\begin{equation}
d\star (v_2 \calf^1) =0 \,,\qquad d\star \left(\frac{v_1^2}{v_2} \calf^2\right) =0\,,
\eqlabel{kpt4}
\end{equation}
resulting in
\begin{equation}
d\star \biggl[\calk\cdot \left(\frac{2\calf^1}{v_1^2}
+\frac{\calf^2}{v_2^2}\right)\biggr]=0\,.
\eqlabel{kpt5}
\end{equation}
However, it is generically violated by Bianchi identities
$d\calf^i=0$. Indeed,
\begin{equation}
d\calf^2= d\left(-\frac{2v_2^2}{v_1^2} \calf^1\right)=-2\ d\left(
\frac{v_2^2}{v_1^2}\right)\wedge \calf^1-2 \frac{v_2^2}{v_1^2}\ d\calf^1\ \ne 0\,,
\eqlabel{kpt6}
\end{equation}
unless $v_i$ are functions of $\{t,r\}$ exclusively
(for purely electric $\calf^1$).
For this reason, the action \eqref{kpt1} does not adequately describe the fluctuations around the background. 

Going back to \eqref{kptf}, we read off the ansatz
\begin{equation}
\calf^1\equiv \frac{Q}{r^2}\ e^{2\eta_1-\frac w2-\frac 32\chi}\ dt\wedge dr\,,\qquad
\calf^2=-2\frac{v_2^2}{v_1^2}\ \calf^1=-2 e^{2(\eta_2-\eta_1)}\ \calf^1\,, 
\eqlabel{kpt9}
\end{equation}
from which we recover from \eqref{kpt1} the effective
one-dimensional Lagrangian of \cite{Klebanov:2010tj},
\begin{equation}
\begin{split}
\call=&\frac{r^2}{L^2}e^{-\frac w2}
\biggl[\frac{63\calg}{8}\chi'^2+\frac \calg2\left(2\eta_1'^2+\eta_2'^2\right)
+\calg(2\eta_1'+\eta_2')^2+\frac {2\calg}{r}w'-\frac 2r \calg'\\
&-\frac{2\calg}{r^2}+V_Q+V_s
\biggr]\,,
\end{split}
\eqlabel{kpt10}
\end{equation}
where
\begin{equation}
\begin{split}
&V_Q=\frac{4L^2}{r^4}e^{-\frac 32 \chi}\left(e^{2\eta_1}+2e^{2\eta_2}\right)\ Q^2\,,\\
&V_s=\frac{9}{2L^2}e^{-\frac {21}{2}\chi}-\frac{4}{L^2}e^{-\frac 92\chi}
\left(2e^{-\eta_1}+e^{-\eta_2}\right)+\frac{1}{2L^2}e^{-2(2\eta_1+\eta_2)
-\frac 92\chi}\left[2e^{-2\eta_1}+e^{-2\eta_2}\right]\,.
\end{split}
\eqlabel{kpt11}
\end{equation}
The Lagrangian $\call$ needs to be supplemented with the zero-energy
constraint, coming from the $rr$-component of the Einstein equations,
\begin{equation}
\frac 2r \calg'-\calg\biggl[\frac{63}{8}\chi'^2+\frac 12\left(2\eta_1'^2+\eta_2'^2\right)
+(2\eta_1'+\eta_2')^2+\frac {2}{r}w'
-\frac{2}{r^2}\biggr]+V_Q+V_s=0\,.
\eqlabel{kpt12}
\end{equation}

As an independent check, we reproduce from \eqref{kpt10}-\eqref{kpt12}
the analytic extremal $AdS_2\times \reals^2$ solution of \cite{Klebanov:2010tj},
\begin{equation}
\begin{split}
&\calg = \mathbf{4} \cdot \frac{2^{\frac 54}(r^4-1)^2}{3^{\frac 74}r^{12}}\,,
\qquad w=w_0-14\ln r\,,\qquad \eta_1=\frac 17\ln 3\\
&\eta_2=\frac 17\ln 3-\ln 2\,,\qquad \chi=\frac{5}{14}\ln 3 -\frac 12\ln 2
+\frac43\ln r \,,\qquad Q = \mathbf{4} \cdot
\frac{2^{\frac 74}}{3^{\frac 54}} \,.
\end{split}
\eqlabel{kpt13}
\end{equation}
The coefficients marked in boldface in \eqref{kpt13} differ from
\cite{Klebanov:2010tj}, owing to the fact that their background had AdS length $L=1$.

\subsection{Background}\label{backpt}

Assuming an equivalent form of the 4D metric and field strengths as in the previous section,
\begin{equation}
ds_4^2=-\frac{4\alpha^2 f}{r^2}\ dt^2+\frac{4\alpha^2}{r^2} d\bm{x}^2
+ \frac{s^2}{4r^2 f}\ dr^2\,,\ \ \calf^1=\frac{q\alpha s}{v_2}\ dr\wedge dt
\,,\ \ \calf^2=-\frac{2v_2^2}{v_1^2}\calf^1\,,
\eqlabel{4dm}
\end{equation}
where $\alpha,q$ are coefficients (related to the temperature and the baryonic
charge), and $f,s,v_i,g\equiv e^\phi$ are all functions of $r$,
we derive the following equations of motion:
\begin{equation}
\begin{split}
&0=f'+f \biggl(
\frac{r v_2'^2}{4v_2^2}+\frac{rv_1'^2}{2v_1^2} +\frac{rg'^2}{g^2}-\frac3r\biggr)
-\frac{s^2 r^3 (2 v_2^2+v_1^2)q^2}{8v_2 v_1^2}
-\frac{s^2 g^4(2 v_2^2 v_1^2+v_1^4+9)}{4v_2 v_1^2 r} 
\\&+\frac{2 g^2 s^2 (2 v_2+v_1)}{v_2 v_1 r}\,,
\end{split}
\eqlabel{eom1}
\end{equation}
\begin{equation}
\begin{split}
0=s'+\frac{s r}{4} \biggl(
\frac{v_2'^2}{v_2^2}+\frac{2 v_1'^2}{v_1^2}+\frac{4 g'^2}{g^2}
\biggr)\,,
\end{split}
\eqlabel{eom2}
\end{equation}
\begin{equation}
\begin{split}
&0=v_1''-\frac{v_1'^2}{v_1}+v_1' \biggl(
\frac{s^2g^4 (2 v_2^2 v_1^2+v_1^4+9)}{4f v_2 v_1^2 r}
-\frac{2 s^2 g^2(2 v_2+v_1)}{v_1 f v_2 r}
+\frac{s^2 r^3 q^2(2 v_2^2+v_1^2)}{8f v_2 v_1^2} +\frac 1r\biggr)
\\&-\frac{s^2g^4 (v_1^4-9)}{2v_1 f v_2 r^2}+\frac{s^2 r^2 v_2 q^2}{2v_1 f}
-\frac{4 s^2 g^2}{f r^2}\,,
\end{split}
\eqlabel{eom3}
\end{equation}
\begin{equation}
\begin{split}
&0=v_2''-\frac{v_2'^2}{v_2}+v_2' \biggl(
\frac{s^2g^4 (2 v_2^2 v_1^2+v_1^4+9)}{4f v_2 v_1^2 r}
-\frac{2 s^2g^2 (2 v_2+v_1)}{v_1 f v_2 r}
+\frac{s^2 r^3 q^2(2 v_2^2+v_1^2)}{8f v_2 v_1^2}+\frac1r\biggr)
\\&-\frac{s^2 (2 v_2^2 v_1^2-v_1^4-9)g^4}{2f v_1^2 r^2}
-\frac{4 s^2g^2}{f r^2} -\frac{s^2 q^2 r^2 (2 v_2^2-v_1^2)}{4f v_1^2}\,,
\end{split}
\eqlabel{eom4}
\end{equation}
\begin{equation}
\begin{split}
&0=g''-\frac{g'^2}{g}+g' \biggl(
\frac{g^4 s^2 (2 v_2^2 v_1^2+v_1^4+9)}{4f v_2 v_1^2 r}
-\frac{2 g^2s^2 (2 v_2+v_1)}{v_1 f v_2 r}
+\frac{s^2 r^3 q^2(2 v_2^2+v_1^2)}{8f v_2 v_1^2}+\frac1r\biggr)\\
&-\frac{s^2 g^5(2 v_2^2 v_1^2+v_1^4+9)}{2f v_2 v_1^2 r^2}
+\frac{2 g^3 s^2 (2 v_2+v_1)}{v_1 f v_2 r^2}\,.
\end{split}
\eqlabel{eom5}
\end{equation}
Following \cite{Cassani:2012pj}, the holographic spectroscopy
relates the scalars $\{v_1,v_2,g\}$ to the boundary gauge theory operators
$\calo_\Delta$ of conformal dimension $\Delta$ as in table \ref{table1}.
\begin{table}[ht]
\caption{Holographic spectroscopy of the background scalars}
\centering
\begin{tabular}{c c c c}
\hline
mass eigenstate & $m^2 L^2$ & $\Delta$ & $U(1)$ R-charge\\
\hline
$\ln [v_1v_2^{-1}]$ & $-2$ & $(2,1)$ & 0 \\
\hline
$\ln [v_1^2v_2 g^3]$ & $4$& $4$ & 0\\
\hline
$\ln [v_1^2v_2 g^{-4}]$ & $18$& $6$ & 0\\
\hline
\end{tabular}
\label{table1}
\end{table}
Notice that the bulk scalar $\ln[v_1 v_2^{-1}]$ can be
identified \cite{Klebanov:1999tb}
either with the operator $\calo_2$, the {\it normal quantization},
or with the operator $\calo_1$, the {\it alternative quantization}.
In \cite{Klebanov:2010tj} the authors consider the normal quantization
only;
here, we discuss both cases.

Eqs.~\eqref{eom1}-\eqref{eom5} should be solved subject to
the following asymptotic expansion
\nxt In the UV, \ie as $r\to 0$, and with the identification
$\ln [v_1v_2^{-1}]\Longleftrightarrow\calo_2$,
we have
\begin{equation}
\begin{split}
&f=1+f_3 r^3+\frac38 q^2 r^4-\frac16 v_{1,2} q^2 r^6+\calo(r^7)\,,\qquad
s=1-\frac32 v_{1,2}^2 r^4+\frac16 v_{1,2} q^2 r^6+\calo(r^7)\,,
\end{split}
\eqlabel{uuv1}
\end{equation}
\begin{equation}
\begin{split}
&v_1=1+v_{1,2} r^2+\biggl(
v_{1,4}+\left(\frac{24}{35} v_{1,2}^2-\frac{1}{35} q^2\right) \ln r\biggr) r^4
-\frac13 f_3 v_{1,2} r^5+\biggl(
v_{1,6}\\&+\biggl(
-\frac{13}{350} v_{1,2} q^2+\frac{156}{175} v_{1,2}^3\biggr) \ln r
\biggr) r^6+\calo(r^7\ln r)\,,
\end{split}
\eqlabel{uuv2}
\end{equation}
\begin{equation}
\begin{split}
&v_2=1-2 v_{1,2} r^2+\biggl(
\frac32 v_{1,2}^2+v_{1,4}+\frac18 q^2
+\biggl(
\frac{24}{35} v_{1,2}^2-\frac{1}{35} q^2\biggr) \ln r\biggr) r^4
+\frac23 f_3 v_{1,2} r^5
+\biggl(
v_{1,6}\\&-\frac{39}{10} v_{1,2} v_{1,4}+\frac{4647}{3500} v_{1,2}^3
-\frac{653}{3500} v_{1,2} q^2+\biggl(
\frac{13}{175} v_{1,2} q^2-\frac{312}{175} v_{1,2}^3\biggr) \ln r\biggr) r^6
+\calo(r^7\ln r )\,,
\end{split}
\eqlabel{uuv3}
\end{equation}
\begin{equation}
\begin{split}
&g=1+\biggl(-\frac{3}{56} v_{1,2}^2+\frac34 v_{1,4}
+\frac{1}{56} q^2+\biggl(
\frac{18}{35} v_{1,2}^2-\frac{3}{140} q^2\biggr) \ln r\biggr) r^4
+\biggl(-v_{1,6}+\frac{13}{10} v_{1,2} v_{1,4}\\&-\frac{1549}{3500} v_{1,2}^3
-\frac{37}{1750} v_{1,2} q^2\biggr) r^6+\calo(r^7\ln r)\,,
\end{split}
\eqlabel{uuv4}
\end{equation}
i.e.~the UV part of the solution is characterized (given $q$) by
\begin{equation}
\biggl\{\ f_3\,,\ v_{1,2}\,,\ v_{1,4}\,,\ v_{1,6}\biggr\}\,;
\eqlabel{uvpar}
\end{equation}
\nxt in the UV, \ie as $r\to 0$, and  instead with the identification
$\ln [v_1v_2^{-1}]\Longleftrightarrow\calo_1$,
we have
\begin{equation}
\begin{split}
&f=1+f_3 r^3+\frac38 q^2 r^4+\biggl(-\frac{9}{20}v_{1,1}^2f_3-\frac{3}{10}v_{1,1}q^2\biggr) r^5+\frac{37}{120}v_{1,1}^2 q^2r^6+\calo(r^7)\,,
\end{split}
\eqlabel{uva1}
\end{equation}
\begin{equation}
\begin{split}
&s=1-\frac34 v_{1,1}^2 r^2+\frac{489}{800} v_{1,1}^4 r^4
+\biggl(
v_{1,1}^5+\frac25 v_{1,1}^2 f_3+\frac{1}{10} v_{1,1} q^2\biggr) r^5
+\biggl(
\frac{5661}{22400} v_{1,1}^6\\
&+\frac18 v_{1,1}^3 f_3-\frac{269}{1680} v_{1,1}^2 q^2
+\frac34 v_{1,1}^2 v_{1,4}+\biggl(-\frac{51}{70} v_{1,1}^6
-\frac{3}{140} v_{1,1}^2 q^2\biggr) \ln r
\biggr) r^6+\calo(r^7\ln r)\,,
\end{split}
\eqlabel{uva2}
\end{equation}
\begin{equation}
\begin{split}
&v_1=1+v_{1,1} r-\frac15 v_{1,1}^2 r^2-\frac{31}{20} v_{1,1}^3 r^3
+\biggl(
v_{1,4}+\biggl(-\frac{34}{35} v_{1,1}^4-\frac{1}{35} q^2\biggr) \ln r\biggr) r^4
+\biggl(-\frac{103}{800} v_{1,1}^5\\
&+\frac{19}{60} v_{1,1}^2 f_3+\frac{11}{120} v_{1,1} q^2+\frac32 v_{1,1} v_{1,4}
+\biggl(
-\frac{3}{70} v_{1,1} q^2-\frac{51}{35} v_{1,1}^5\biggr) \ln r\biggr) r^5
+\biggl(
v_{1,6}\\&+\biggl(
-\frac{51}{70} v_{1,1}^6-\frac{3}{140} v_{1,1}^2 q^2\biggr) \ln r\biggr) r^6+\calo(r^7\ln r)\,,
\end{split}
\eqlabel{uva3}
\end{equation}
\begin{equation}
\begin{split}
&v_2=1-2 v_{1,1} r+\frac{13}{10} v_{1,1}^2 r^2+\frac{1}{10} v_{1,1}^3 r^3
+\biggl(
\frac{131}{40} v_{1,1}^4+\frac12 v_{1,1} f_3+v_{1,4}+\frac18 q^2
+\biggl(-\frac{34}{35} v_{1,1}^4\\&-\frac{1}{35} q^2\biggr) \ln r\biggr) r^4
+\biggl(
-\frac{4597}{400} v_{1,1}^5-\frac{14}{15} v_{1,1}^2 f_3-\frac{13}{30} v_{1,1} q^2
-3 v_{1,1} v_{1,4}+\biggl(
\frac{3}{35} v_{1,1} q^2\\&
+\frac{102}{35} v_{1,1}^5\biggr) \ln r\biggr) r^5
+\biggl(
\frac{166743}{14000} v_{1,1}^6-\frac{29}{40} v_{1,1}^3 f_3
+\frac{8061}{14000} v_{1,1}^2 q^2+\frac{39}{20} v_{1,1}^2 v_{1,4}+v_{1,6}
\\&+\biggl(-\frac{459}{175} v_{1,1}^6-\frac{27}{350} v_{1,1}^2 q^2\biggr) \ln r\biggr) r^6
+\calo(r^7\ln r)\,,
\end{split}
\eqlabel{uva4}
\end{equation}
\begin{equation}
\begin{split}
&g=1-\frac{3}{10} v_{1,1}^2 r^2-\frac12 v_{1,1}^3 r^3+\biggl(
\frac{2047}{1400} v_{1,1}^4+\frac18 v_{1,1} f_3+\frac34 v_{1,4}
+\frac{1}{56} q^2+\biggl(-\frac{51}{70} v_{1,1}^4\\&-\frac{3}{140} q^2\biggr) \ln r\biggr) r^4
+\biggl(
-\frac{73}{40} v_{1,1}^5+\frac{1}{10} v_{1,1}^2 f_3\biggr) r^5
+\biggl(
-\frac{6761}{14000} v_{1,1}^6+\frac{283}{240} v_{1,1}^3 f_3
\\&+\frac{2837}{21000} v_{1,1}^2 q^2
+\frac{19}{40} v_{1,1}^2 v_{1,4}-v_{1,6}+\biggl(
\frac{187}{700} v_{1,1}^6+\frac{11}{1400} v_{1,1}^2 q^2\biggr) \ln r\biggr) r^6+\calo(r^7\ln r)\,,
\end{split}
\eqlabel{uva5}
\end{equation}
characterized  (given $q$) by
\begin{equation}
\biggl\{\ v_{1,1}\,,\ f_3\,,\  v_{1,4}\,,\ v_{1,6}\biggr\}\,;
\eqlabel{uvpara}
\end{equation}
\nxt in the IR, \ie as $y\equiv 1-r\to 0$, we have
\begin{equation}
\begin{split}
&f=-\frac{(s^h_0)^2}{8v^h_{2,0} (v^h_{1,0})^2} \biggl(
2 (g^h_0)^4 \biggl((v^h_{1,0})^4+2 (v^h_{1,0})^2 (v^h_{2,0})^2+9\biggr)
-16 (g^h_0)^2 v^h_{1,0} \biggl(v^h_{1,0}+2 v^h_{2,0}\biggr)
\\&+q^2 \biggl((v^h_{1,0})^2+2 (v^h_{2,0})^2\biggr)
\biggr) y+\calo(y^2)\,,\\
&s=s^h_0+\calo(y)\,,\qquad v_i=v^h_{i,0}+\calo(y)\,,\qquad g=g^h_0+\calo(y)\,,
\end{split}
\eqlabel{ir}
\end{equation}
characterized  (given $q$) by
\begin{equation}
\biggl\{\ s^h_0\,,\ v_{1,0}^h\,,\ v_{2,0}^h\,,\ g^h_0\biggr\}\,.
\eqlabel{irpar}
\end{equation}

\begin{figure}[t]
\begin{center}
\psfrag{q}{{$q/q_{crit}$}}
\psfrag{v}[b]{{$v_{1,0}^h/v_1^{crit}-1$}}
\psfrag{u}[t]{{$v_{2,0}^h/v_2^{crit}-1$}}
  \includegraphics[width=3in]{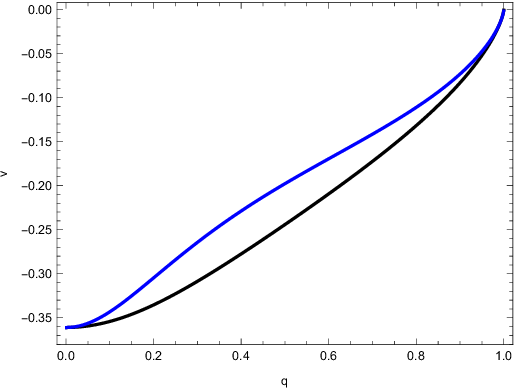}
  \includegraphics[width=3in]{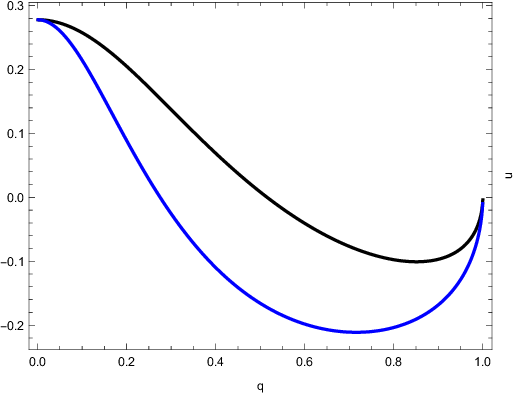}
\end{center}
 \caption{The values of the bulk scalars
 $v_1$ and $v_2$ at the horizon for different
 quantizations of the mode $\ln[v_1v_2^{-1}]$ (see
 table \ref{table1}): $\calo_2$ (black) and $\calo_1$ (blue).
 The limit $q/q_{crit}\to 1$ is a quantum critical regime corresponding to
 the zero-temperature limit $T\to 0$ of the baryonic black membranes \eqref{sol2} and \eqref{so2ir};
  the regime $q/q_{crit}\to 0$ is the black membrane solution with vanishing
  baryonic charge density --- the $AdS_4$-Schwarzschild background
  with the trivial profile for the scalars $v_1=v_2\equiv 1$
  \eqref{sol1}.
}\label{figure1}
\end{figure}

\begin{figure}[t]
\begin{center}
\psfrag{q}{{$q/q_{crit}$}}
\psfrag{g}[b]{{$g_{0}^h/g^{crit}-1$}}
\psfrag{t}[t]{{$T/\alpha$}}
  \includegraphics[width=3in]{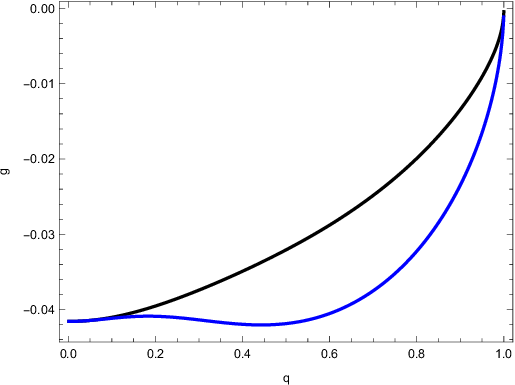}
  \includegraphics[width=3in]{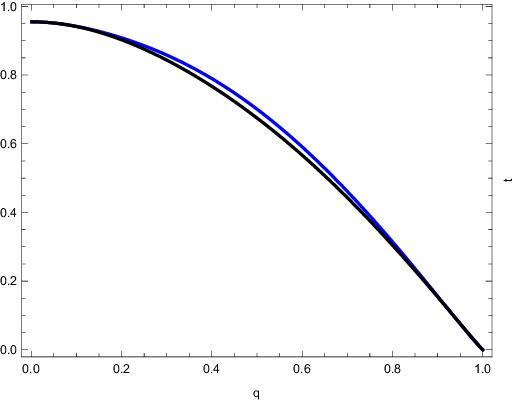}
\end{center}
 \caption{The values of the bulk scalar
 $g$  at the horizon and the reduced
temperature $T/\alpha$ (see \eqref{tbaryonic})
for different
 quantizations of the mode $\ln[v_1v_2^{-1}]$ (see
 table \ref{table1}): $\calo_2$ (black) and $\calo_1$ (blue).
 The limit $q/q_{crit}\to 1$ is a quantum critical regime corresponding to
 $T\to 0$ of the baryonic black membranes \eqref{sol2} and \eqref{so2ir};
  the regime $q/q_{crit}\to 0$ is the black membrane solution with vanishing
  baryonic charge density --- the $AdS_4$-Schwarzschild background
  with the trivial profile for the scalar $g\equiv 1$
  \eqref{sol1}.
}\label{figure2}
\end{figure}

There are two exact analytic solutions of \eqref{eom1}-\eqref{eom5}:
\begin{itemize}
\item an $AdS_4$-Schwarzschild solution,
\begin{equation}
q=0:\qquad f=1-r^3\,,\qquad s=v_i=g=1\,;
\eqlabel{sol1}
\end{equation}
\item the extremal  $AdS_2\times \reals^2$ solution of
\cite{Klebanov:2010tj} given in \eqref{kpt13}
\begin{equation}
q=q_{crit}=\frac{2^{\frac{15}{4}}}{3^{\frac 54}}:\qquad f=\frac{32(s^h_0)^2\ (1-r^4)^2}
{27q_{crit}\ r^8}\,,\ s=\frac{s^h_0}{r^7}\,,\
v_i=\frac{v_i^{crit}}{r^2}\,,\ g=g^{crit} r^2\,,
\eqlabel{sol2}
\end{equation}
where
\begin{equation}
v_1^{crit}=\frac{16}{3q_{crit}}\,,\qquad
v_2^{crit}=\frac{8}{3q_{crit}}\,,\qquad g^{crit}=\frac{32}{9q_{crit}} \,.
\eqlabel{so2ir}
\end{equation}
\end{itemize}

A baryonic black membrane solution is an interpolating solution
for $q\in [0,q_{crit}]$.
We use numerical technique developed in \cite{Aharony:2007vg}
to construct such solutions. Some results of such computations,
which validate \eqref{sol2}, 
are presented in figs.~\ref{figure1}-\ref{figure2}.

Given $q$, a numerical solution is
characterized by \eqref{uvpar} (or \eqref{uvpara}) and \eqref{irpar},
which determine the black membrane Hawking temperature $T$,
\begin{equation}
\begin{split}
&\frac T\alpha=\frac{s^h_0}{8 \pi v^h_{2,0} (v^h_{1,0})^2} \biggl(
2 (g^h_0)^2 \biggl(
8 v^h_{1,0} (v^h_{1,0}+2 v^h_{2,0})-(g^h_0)^2 ((v^h_{1,0})^4
+2 (v^h_{1,0})^2 (v^h_{2,0})^2+9)\biggr)
\\&-\biggl((v^h_{1,0})^2+2 (v^h_{2,0})^2\biggr)q^2\biggr)\,.
\end{split}
\eqlabel{tbaryonic}
\end{equation}
From \eqref{sol2} and \eqref{so2ir}, as $q\to q_{crit}$, we have from
\eqref{tbaryonic} (see also the right panel of fig.~\ref{figure2})
\begin{equation}
\lim_{q\to q_{crit}} \frac{T}{\alpha}\ \propto\  \lim \left(1-\frac{q}{q_{crit}}
\right)\ =0\,.
\eqlabel{extt}
\end{equation}

\subsection{Fluctuations}\label{fluckpt}

Given these backgrounds, we now discuss fluctuations of the $R$-charge density. 
To this end, we need to supplement the Lagrangian $\call_{KPT}$ of
\eqref{kpt1} to quadratic
order in $\{\ \calf^0\,,\ B\,,\ b_i\ \}$ obtained from the full
effective action $\call_{CKV}$
\begin{equation}
\begin{split}
&\call_{KPT}\to \call_{KPT}+\delta\call_2[\calf^0,B,b_i]\equiv \call_{KPT}+
\delta\call_{2,kin}-\delta V_2\star 1\,,\\
&\delta\call_{2,kin}=-\frac 14 v_i^{-2}(\del b_i)^2\star 1-\frac 14 K_{ijk}b_k \calf^i\wedge \calf^j
-\frac{\calk}{4} \calf^0\wedge \star \calf^0
+\frac{\calk}{2}\calf^0\wedge \star \sum_i \frac{b_i}{v_i^{2}} \calf^i
\\&-\frac 14 e^{-4\phi} dB\wedge \star dB
-\calk \sum_iv_i^{-2}\ B\wedge\star B+B\wedge\biggl\{
\star \left(\sum_i \calk v_i^{-2} \calf^i\right) +\frac 12\calk_{ijk}b_km_i \calf^j\\
&+\frac {e_0}{2} \calf^0 
\biggr\}\,,\\
&\delta V_2={e^{4\phi}}\calk^{-1}\biggl(2v_1^2(b_1+b_2)^2+4 b_1^2 v_2^2\biggr) 
+ e_0 e^{4\phi}\calk^{-1}\biggl(2 b_1b_2+b_1^2\biggr)\,. 
\end{split}
\eqlabel{dl}
\end{equation}
Dualizing $B$ as per \eqref{baction} and \eqref{ahaction} we get 
\begin{equation}
\begin{split}
&\delta\call_{2,kin}=-\frac 14 v_i^{-2}(\del b_i)^2\star 1-\frac 14 K_{ijk}b_k \calf^i\wedge \calf^j
-\frac{\calk}{4} \calf^0\wedge \star \calf^0
+\frac{\calk}{2}\calf^0\wedge \star \sum_i \frac{b_i}{v_i^{2}} \calf^i\\
&-\frac{1}{M^2} \left(dA_H-\frac 12 J_2\right)\wedge\star \left(dA_H
-\frac 12 J_2\right)-4 e^{4\phi} A_H\wedge\star A_H\,,
\end{split}
\eqlabel{dlk}
\end{equation}
with
\begin{equation}
\begin{split}
&M^2=2v_2+\frac{v_1^2}{v_2}\,,\qquad J_a={-}2(b_1+b_2)\calf^1 {-} 2b_1 \calf^2
-\frac{e_0}{2} \calf^0\,,\qquad J_b= {-}2v_2 \calf^1 {-}
\frac{v_1^2}{v_2}\calf^2\,.
\end{split}
\eqlabel{defmjajb}
\end{equation}

Within the effective action \eqref{dl} we consider linearized
fluctuations
\begin{align}
A^0 &= \delta A^0_t\ dt+\delta A^0_{x_2}\ d x_2+\delta A^0_r\ dr\,,&
A_H &= \delta A^H_t\ dt+\delta A^H_{x_2}\ d x_2+\delta A^H_r\ dr\,, \nonumber \\
\delta A^i &=\delta A^i_{x_1}\ dx_1+\delta A^i_r\ dr\,,&
b_i &= \delta b_i\,,
\eqlabel{fldec}
\end{align}
about the black membrane background \eqref{4dm}, 
which we take to be functions of $t$, $r$, and $x_2$ as follows:
\begin{equation}
A^{0,H}_{t,x_2,r}=e^{-i w t +i k x_2}\cdot  \cala^{0,H}_{t,2,r}(r) \quad
\delta A^{i}_{x_1,r}=e^{-i w t +i k x_2}\cdot \cala^{i}_{1,r}(r)\,,\quad
\delta b_i= e^{-i w t +i k x_2}\cdot \calb_i(r) \,.
\end{equation}
Note that $\delta A^i$ is turned on “magnetically”, in the sense that it is the only form with support in the $x_1$ direction, while we are considering fluctuations with a spatial profile in the $x_2$ direction.
It is straightforward
to verify that the set \eqref{fldec} will decouple from the remaining
fluctuations in the helicity-0 (the sound channel) sector.
We use the bulk gauge transformations \eqref{gauge} to set
\begin{equation}
\delta \cala^i_r=\delta \cala^0_r\equiv 0\,.
\eqlabel{fixgauge}
\end{equation}
which lead to the constraints
\begin{equation}
\begin{split}
&0=(\cala^H_2)'+\frac{c_2^2 w}{c_1^2 k} (\cala^H_t)'
+\frac32 (\cala^0_2)'+\frac{3c_2^2 w}{2c_1^2 k}  (\cala^0_t)'
{-}\frac{c_2^2 w (\calb_1 (2 v_2^2-v_1^2)-\calb_2 v_1^2) F}{c_1^2 v_1^2 k}
\\&-\frac{i}{c_1^2 v_2 k}
\biggl(4 c_1^2  (2 v_2^2+  v_1^2  ) c_2^2 g^4-v_2 (c_2^2 w^2-c_1^2 k^2)\biggr)
\cala^H_r\,,
\end{split}
\eqlabel{fl8}
\end{equation}
\begin{equation}
\begin{split}
&0=(2 v_2^2 v_1^2+v_1^4+9) \biggl(
(\cala^0_t)'+\frac{c_1^2 k}{c_2^2 w} (\cala^0_2)'
\biggr)
+6 (\cala^H_t)'
+\frac{6 c_1^2 k}{c_2^2 w} (\cala^H_2)'+F \biggl(
-\frac{2}{v_1^2} (2 v_2^2 v_1^2\\&+v_1^4 {+}6 v_2^2 {-} 3 v_1^2) \calb_1
+2 \calb_2 (2 v_2^2+v_1^2 {+}3)\biggr)
- \frac{6 i}w  \frac{(c_1^2 k^2-c_2^2 w^2)}{c_2^2}  \cala^H_r \,,
\end{split}
\eqlabel{fl9}
\end{equation}
as well as 2 more equations which can be solved for the metric components.
The equations of motion for the remaining fluctuations
take the form
\begin{equation}
\begin{split}
&0=(\cala^0_t)''+\biggl(
\frac{2 v_1'}{v_1}-\frac{c_3'}{c_3}+\frac{2c_2'}{c_2}-\frac{c_1'}{c_1}
+\frac{v_2'}{v_2}\biggr) (\cala^0_t)'
-\frac{c_3^2 k}{c_2^2} (\cala^0_t k+\cala^0_2 w)
+\frac{24 g^4 c_3^2}{v_2 v_1^2} \cala^H_t\\&-\frac{2 F'}{v_1^2} (\calb_1-\calb_2)
+F \biggl(
-\frac{2}{v_1^2} (\calb_1'-\calb_2')
+\biggl(
-\frac{2v_2'}{v_2 v_1^2}+\frac{2c_3'}{v_1^2 c_3}
-\frac{4 c_2'}{c_2 v_1^2}+\frac{2c_1'}{v_1^2 c_1}
\biggr) (\calb_1-\calb_2)\biggr)\,,
\end{split}
\eqlabel{fl1}
\end{equation}
\begin{equation}
\begin{split}
&0=(\cala^0_2)''+\biggl(
\frac{2 v_1'}{v_1}+\frac{v_2'}{v_2}-\frac{c_3'}{c_3}
+\frac{c_1'}{c_1}\biggr) (\cala^0_2)'
+\frac{c_3^2 w}{c_1^2} (\cala^0_t k+\cala^0_2 w)+\frac{24 g^4  c_3^2}{v_2 v_1^2}\cala^H_2\,,
\end{split}
\eqlabel{fl2}
\end{equation}
\begin{equation}
\begin{split}
&0=(\cala^H_t)''+\biggl(
\frac{2 c_2'}{c_2}-\frac{c_1'}{c_1}-\frac{c_3'}{c_3}
+ \frac{v_1^2-2 v_2^2}{(2 v_2^2+v_1^2) v_2} v_2'
-  \frac{2 v_1}{2 v_2^2+v_1^2} v_1'\biggr) (\cala^H_t)'
-\biggl(4 g^4 c_2^2 (2 v_2^2 v_1^2\\& + v_1^4 + 9 ) + v_2 v_1^2 k^2\biggr)
\frac{c_3^2}{v_1^2 c_2^2 v_2} \cala^H_t
+i w (\cala^H_r)'+i w \cala^H_r \biggl(
\frac{2 c_2'}{c_2}-\frac{c_1'}{c_1}-\frac{c_3'}{c_3} -  \frac{2 v_1}{2 v_2^2+v_1^2} v_1'
\\&+  \frac{v_1^2-2 v_2^2}{(2 v_2^2+v_1^2) v_2} v_2'\biggr)
-\frac{6 (v_1 v_2 v_2'  + v_1' ( v_1^2+v_2^2))}{v_1 (2 v_2^2+v_1^2)} (\cala^0_t)'
+\frac{( v_1^2 -2 v_2^2+3) F}{v_1^2} \calb_1'\\&+\frac{(v_1^2-3) F}{v_1^2} \calb_2'
+ \frac{2 i c_1 c_3 v_1 k (v_2' v_1-v_2 v_1')}{(2 v_2^2+v_1^2) c_2^2} \biggl(\cala^1_1-\cala^2_1\biggr)
+\biggl(
\frac{(v_1^2-2 v_2^2+3) F}{v_1^2} \biggl(
\frac{2 c_2'}{c_2}-\frac{c_1'}{c_1}\\&-\frac{c_3'}{c_3}+\frac{F'}{F}\biggr)
-\frac{2 F (v_1^4-4 v_2^2 v_1^2-4 v_2^4)}{(2 v_2^2+v_1^2) v_1^3} v_1'
+ v_2' \frac{F(v_1^4-8 v_2^2 v_1^2-4 v_2^4+3 v_1^2+6 v_2^2)}{v_1^2 (2 v_2^2+v_1^2) v_2}
\biggr) \calb_1
\\&-\frac{c_3^2 k w}{c_2^2} \cala^H_2+\biggl(
\frac{(v_1^2-3) F}{v_1^2} \biggl(
\frac{2 c_2'}{c_2}-\frac{c_1'}{c_1}-\frac{c_3'}{c_3}+\frac{F'}{F}\biggr)
-\frac{2 F v_1}{2 v_2^2+v_1^2} v_1'\\&+\frac{F (v_1^4-2 v_2^2 v_1^2-3 v_1^2-6 v_2^2)}{v_1^2 (2 v_2^2+v_1^2) v_2} v_2'
\biggr) \calb_2\,,
\end{split}
\eqlabel{fl3}
\end{equation}
\begin{equation}
\begin{split}
&0=(\cala^H_2)''+\biggl(
\frac{c_1'}{c_1}-\frac{c_3'}{c_3}-  \frac{2 v_1}{2 v_2^2+v_1^2} v_1'
-\frac{2 v_2^2-v_1^2}{v_2 (2 v_2^2+v_1^2)} v_2'\biggr) (\cala^H_2)'
-\biggl(4 c_1^2 g^4 ( 2 v_2^2 v_1^2 \\&+  v_1^4+9)-v_2 v_1^2 w^2\biggr) \frac{c_3^2}{v_2 v_1^2 c_1^2} \cala^H_2
+\frac{c_3^2 k w}{c_1^2} \cala^H_t-i k (\cala^H_r)'
\\&+i k \biggl(
\frac{c_3'}{c_3}-\frac{c_1'}{c_1}+  \frac{2 v_1}{2 v_2^2+v_1^2} v_1'+  \frac{2 v_2^2-v_1^2}{v_2 (2 v_2^2+v_1^2)} v_2'\biggr) \cala^H_r
\\&
-\frac{6 ( v_1 v_2 v_2'  + v_1' ( v_1^2+v_2^2)}
{v_1 (2 v_2^2+v_1^2)} (\cala^0_2)'
-\frac{2 i c_3 v_1 w (v_2 v_1'-v_1 v_2')}{c_1 (2 v_2^2+v_1^2)}
\biggl(\cala^2_1-\cala^1_1\biggr)\,,
\end{split}
\eqlabel{fl4}
\end{equation}
\begin{equation}
\begin{split}
&0=\biggl(\cala^1_1-\cala^2_1\biggr)''+\biggl(
\frac{c_1'}{c_1}-\frac{c_3'}{c_3}+\frac{4 v_2^2}{v_1 (2 v_2^2+v_1^2)}
v_1'-\frac{2 v_2^2-v_1^2}{v_2 (2 v_2^2+v_1^2)} v_2'
\biggr) \biggl(\cala^1_1-\cala^2_1\biggr)'
\\&-\frac{2 i (v_2 v_1'-v_1 v_2') c_3}
{c_1 v_1 (2 v_2^2+v_1^2)} \biggl(
2 k \cala^H_t+2 w \cala^H_2+3 k \cala^0_t+3 w \cala^0_2\biggr)
+\frac{i F k c_3}{c_1 v_2}
\biggl(\calb_2-\frac{4 v_2^2}{v_1^2} \calb_1\biggr)
\\&-\frac{(c_2^2 w^2-c_1^2 k^2) c_3^2}{c_1^2 c_2^2} \biggl(
\cala^2_1-\cala^1_1\biggr)\,,
\end{split}
\eqlabel{fl5}
\end{equation}
\begin{equation}
\begin{split}
&0=(\calb_1)''+\biggl(
\frac{2 c_2'}{c_2}+\frac{c_1'}{c_1}-\frac{c_3'}{c_3}
-\frac{2 v_1'}{v_1}\biggr) (\calb_1)'
+\biggl(-4 c_2^2 c_1^2 c_3^2 v_1^2 (2 v_2^2+v_1^2+3) (2 v_2^2+v_1^2) g^4
\\&+c_3^2 v_2 v_1^2 (2 v_2^2+v_1^2) (c_2^2 w^2-c_1^2 k^2)
+2 c_2^2 F^2 v_2^2 (2 v_2^2-v_1^2)^2 \biggr)
\biggl(c_1^2 v_1^2 v_2 (2 v_2^2+v_1^2) c_2^2\biggr)^{-1} \calb_1
\\&-\frac{F v_2 (2 v_2^2 v_1^2+v_1^4+6 v_2^2-3 v_1^2  )}{c_1^2 (2 v_2^2+v_1^2)}
(\cala^0_t)'-\frac{2 F v_2 (2 v_2^2-v_1^2)}{c_1^2 (2 v_2^2+v_1^2)}
(\cala^H_t)'-2 \biggl(
2 c_1^2 c_3^2 (v_1^2+3) (2 v_2^2\\&+v_1^2) g^4 + F^2 v_2^2 (2 v_2^2-v_1^2)\biggr)
\biggl(c_1^2 v_2 (2 v_2^2+v_1^2)\biggr) \calb_2
-\frac{2 i F v_2 w (2 v_2^2-v_1^2)}{c_1^2 (2 v_2^2+v_1^2)} \cala^H_r
\\&-\frac{4 i v_1^2 c_3 F v_2^2 k}{c_1 (2 v_2^2+v_1^2) c_2^2}
\biggl(\cala^2_1-\cala^1_1\biggr)\,,
\end{split}
\eqlabel{fl6}
\end{equation}
\begin{equation}
\begin{split}
&0=(\calb_2)''+\biggl(
\frac{2 c_2'}{c_2}+\frac{c_1'}{c_1}-\frac{c_3'}{c_3}-\frac{2v_2'}{v_2}\biggr)
(\calb_2)'
+\biggl(-8 c_2^2 c_1^2 c_3^2 v_2 (2 v_2^2+v_1^2) g^4+c_3^2 (2 v_2^2\\
&+v_1^2) (c_2^2 w^2-c_1^2 k^2)+4 c_2^2 F^2 v_2^3\biggr)
\biggl(c_1^2 (2 v_2^2+v_1^2) c_2^2\biggr)^{-1} \calb_2
+\frac{2 F v_2^3 (2 v_2^2+v_1^2+3)}{c_1^2 (2 v_2^2+v_1^2)} (\cala^0_t)'
\\&+\frac{4 F v_2^3}{c_1^2 (2 v_2^2+v_1^2)} (\cala^H_t)'-4 \biggl(
2 c_1^2 c_3^2 (v_1^2+3) (2 v_2^2+v_1^2) g^4+F^2 v_2^2 (2 v_2^2-v_1^2)\biggr) v_2
\biggr(c_1^2 v_1^2 (2 v_2^2\\&+v_1^2)\biggr)^{-1} \calb_1
+\frac{4 i v_2^3 F w}{c_1^2 (2 v_2^2+v_1^2)} \cala^H_r
+\frac{2 i v_2^2 c_3 F v_1^2 k}{c_1 (2 v_2^2+v_1^2) c_2^2} \biggl(
\cala^2_1-\cala^1_1\biggr)\,,
\end{split}
\eqlabel{fl7}
\end{equation}
where, compare with \eqref{4dm},
\begin{equation}
c_1=\frac{2\alpha \sqrt f}{r}\,,\qquad
c_2=\frac{2\alpha}{r}\,,\qquad c_3=\frac{s}{2r\sqrt f }\,,\qquad
F=\frac{q\alpha s}{v_2}\,.
\eqlabel{defcf}
\end{equation}
We explicitly verified that \eqref{fl8} and \eqref{fl9} are consistent with
\eqref{fl1}-\eqref{fl7}.
Fluctuations of the $U(1)$ $R$-charge bulk potential $A^0$ excite
the axions $b_1$ and $b_2$ (see \eqref{fl6} and \eqref{fl7}).
Following \cite{Cassani:2012pj}, the holographic spectroscopy relates
the pseudoscalars $\{b_1,b_2\}$ to the boundary gauge theory operators
$\delta\calo_\Delta$ of conformal dimension $\Delta$ as in table \ref{table2}.
\begin{table}[ht]
\caption{Holographic spectroscopy of the pseudoscalars}
\centering
\begin{tabular}{c c c c}
\hline
mass eigenstate & $m^2 L^2$ & $\Delta$ & $U(1)$ R-charge\\
\hline
$b_1-b_2$ & $-2$ & $(2,1)$ & 0 \\
\hline
$2 b_1+b_2$ & $10$& $5$ & 0\\
\hline
\end{tabular}
\label{table2}
\end{table}
Here again, we have the choice to quantize one of the fluctuations so that it corresponds either to a CFT operator of dimension 2 (\textit{normal quantization}) or of dimension 1 (\textit{alternative quantization}). This choice is independent from the choice of quantization for the background solution.

To proceed we introduce
\begin{equation}
Z\equiv \kk\ \cala^0_t + \ww\ \cala^0_2\,,\qquad \cala\equiv
\cala^1_1-\cala^2_1\,,
\end{equation}
where
\begin{equation}
\ww=\frac{w}{2\pi T}\,,\qquad \kk=\frac{k}{2\pi T}\,,
\eqlabel{defwk}
\end{equation}
and $T$ is the Hawking temperature of the black membrane. 
We use the constraints \eqref{fl8} and \eqref{fl9} to eliminate
$\cala^H_r$ and obtain from \eqref{fl1}-\eqref{fl7} a decoupled
set of the second-order equations for
\begin{equation}
\{\ Z\,,\ \cala^H_t\,,\ \cala^H_2\,,\ \cala\,,\ \calb_1\,,\ \calb_2\ \}\,.
\eqlabel{set}
\end{equation}
Solutions of the resulting equations with appropriate boundary conditions
determine the spectrum of baryonic black membranes quasinormal 
modes --- equivalently
the physical spectrum of linearized fluctuations in membrane gauge theory plasma with
a baryonic chemical potential. Following \cite{Kovtun:2005ev,Son:2002sd}
we impose the incoming-wave boundary conditions at the black membrane
horizon, and 'normalizability'
at asymptotic $AdS_4$ boundary.
Focusing on the $\Re[\ww]=0$  diffusive branch, and
introducing
\begin{equation}
\begin{split}
&Z=\left(1-r\right)^{-i\ww/2}\ z\,,\qquad \cala^H_t=(1-r)^{-i\ww/2}\ a^H_t\,,\qquad
\cala^H_2=i (1-r)^{-i\ww/2}\ a^H_2\,,\\
&\cala=i (1-r)^{-i\ww/2}\ a\,,\qquad \calb_i=(1-r)^{-i\ww/2}\ B_i\,,\qquad  
\ww=-i v\ \kk\,,
\end{split}
\eqlabel{income}
\end{equation}
we solve the fluctuation equations  subject to the asymptotics: 
\nxt in the UV, \ie as $r\to 0_+$, and with the
identifications\footnote{Likewise, we develop the UV expansions for the
alternative quantization of either the background, $\ln[v_1 v_2^{-1}]$, or the fluctuation, $(b_1-b_2)$, (pseudo)scalars: $\{\calo_2,\delta\calo^b_1\}$, $\{\calo_1,\delta\calo_2^b\}$, and $\{\calo_1,\delta\calo^b_1\}$.}
$\ln [v_1v_2^{-1}]\Longleftrightarrow\calo_2$ and $(b_1-b_2)\Longleftrightarrow\delta\calo_2^b$, 
\begin{equation}
\begin{split}
&z=\kk r -\frac 12 \kk^2 v r^2+\calo(r^3)\,,\quad
a^H_t=2 q b_{1,2} r^3+\left(
a^h_{t,4}-\frac 27 a_1 \kk T \pi v_{1,2} \ln r\right) r^4+\calo(r^5\ln r)\,,\\
&a^H_2=\left(a^h_{2,4}-\frac27 T\pi v_{1,2}\kk v a_1 \ln r\right) r^4+\calo(r^5\ln r)
\,,\quad a=a_1 r-\frac12 a_1\kk  v r^2+\calo(r^3)\,,\\
&B_1=b_{1,2} r^2-\frac12 b_{1,2} \kk v r^3+\frac{1}{24(v^2+1)} \biggl(
\pi^2 T^2 b_{1,2} \kk^2 (v^2+1)^2+3 b_{1,2} (v^2+1) (\kk^2 v^2-2 \kk v\\
&-8 v_{1,2})+2 q\biggr)
r^4+\biggl(b_{1,5}+\frac{1}{84} a_1 \kk T \pi q \ln r\biggr) r^5
+\calo(r^6\ln r)\,,\\
&B_2=-2 b_{1,2} r^2+b_{1,2} \kk r^3 v+\calo(r^4)\,,
\end{split}
\eqlabel{uv1}
\end{equation}
specified, for a fixed background and a momentum $\kk$,  by
\begin{equation}
\biggl\{\
v\,,\ a^h_{t,4}\,,\ a^h_{2,4}\,,\ a_1\,,\ b_{1,2}\,,\ b_{1,5}
\
\biggr\}\,;
\eqlabel{fluv}
\end{equation}
\nxt in the IR, \ie as $y\equiv 1-r\to 0_+$,
\begin{equation}
\begin{split}
&z=z^h_0+\calo(y)\,,\qquad a^{H}_{t,2}=a^{H,h}_{t,2;0}+\calo(y)\,,\qquad a=a^h_0+\calo(y)\,,\qquad
B_i=b^h_{i;0}+\calo(y)\,,
\end{split}
\eqlabel{ir1}
\end{equation}
specified  by
\begin{equation}
\biggl\{\
z^h_0\,,\ a^{H,h}_{t;0}\,,\ a^{H,h}_{2;0}\,,\ a^{h}_{0}\,,\
b^{h}_{1;0}\,,\ b^{h}_{2;0} 
\
\biggr\}\,.
\eqlabel{flir}
\end{equation}
Note that in total we have $6+6=12$ parameters, see \eqref{fluv} and \eqref{flir},
which is precisely what is necessary to identify a solution of a coupled system of
6 second-order ODEs for $\{z, a^H_t, a^H_2, a, B_1, B_2\}$. Furthermore, since the equations are linear in the fluctuations, we can, without loss of generality, normalize the solutions so that
\begin{equation}
\lim_{r\to 0}\ \frac{dz}{dr}=\kk\,.
\eqlabel{normaliz}
\end{equation}

Once we fix the background, and solve the fluctuation equations of motion,
we obtain $v=v(\kk)$. Given $v$ we extract the $R$-charge
diffusion coefficient $\cald$, as 
\begin{equation}
\ww =-i\cdot\ \underbrace{2\pi T D}_{\equiv \cald}\cdot\ \kk^2+\calo(\kk^3)\,,\qquad 
\cald\equiv \frac{dv}{d\kk}\bigg|_{\kk=0}\,.
\eqlabel{defd}
\end{equation}

For general values of $q$ we have to solve the fluctuation equations
numerically. At $q=0$, an analytic solution is possible in the
limit $\kk\to 0$ --- which is precisely what is needed to extract $\cald$,
see \eqref{defd}. Specifically, at $q=0$, we
have\footnote{The background geometry is given by \eqref{sol1}.}
$a^H_t=a^H_2=a=B_1=B_2=0$ and 
\begin{equation}
\begin{split}
&0=z''+\frac{(\kk (r^2+r+1) (r^3-v^2-1)+3 r^2 v) v}{(r^3-v^2-1) (1-r^3)}\
z'-\frac{\kk}{4(1-r^3) (r^3-v^2-1) (r^2+r+1)}\\
&\times \biggl(
(r^3 v^2+3 r^2 v^2+6 r v^2+9 r^2+8 v^2+9 r+9) (r^3-v^2-1) \kk
+2 v (r^2+r+1) (r^4\\
&+2 r^3+2 r v^2+3 r^2+v^2+2 r+1)\biggr)\ z\,.
\end{split}
\eqlabel{an1}
\end{equation}
Furthermore, in the hydrodynamics limit,
\begin{equation}
z=\kk\ r + \kk^3\ z_2(r)+\calo(\kk^5)\,,\qquad v=\cald\ \kk+\calo(\kk^3)\,,
\eqlabel{an2}
\end{equation}
we find (imposing the UV boundary condition \eqref{uv1})
\begin{equation}
\begin{split}
&z_2=-\frac{1}{12} (2 \cald-3) (2 \cald-3 r+3)\ \ln(1-r)
+\frac38 (r+2)\ \ln(r^2+r+1)\\
&+\frac{\sqrt 3}{12}(4 \cald^2-9 r)\  \arctan{\frac{2 r+1}{\sqrt 3}}
+\frac18 r (\sqrt 3 \pi-8 \cald^2)
-\frac{\sqrt 3}{18}  \pi \cald^2+\frac16 \cald^2\  \ln(r^2+r+1)\,.
\end{split}
\eqlabel{z2sol}
\end{equation}
Finally, the regularity of $z_2$ at the horizon, \ie as $r\to 1$, determines
\begin{equation}
\cald\bigg|_{q=0}=\frac 32\,.
\eqlabel{an3}
\end{equation}

\begin{figure}[t]
\begin{center}
\psfrag{d}{{$\cald_R$}}
\psfrag{q}{{$q/q_{crit}$}}
  \includegraphics[width=4in]{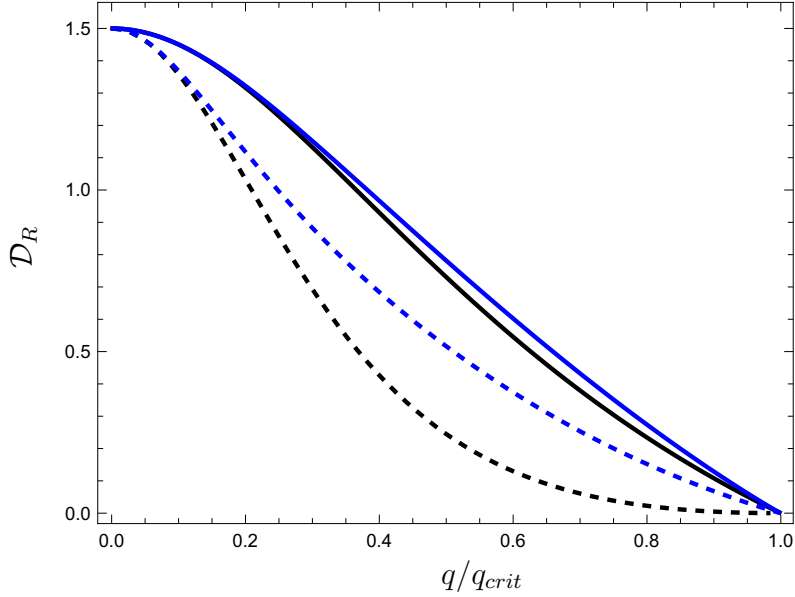}
\end{center}
 \caption{$R$-charge dimensionless diffusion coefficient
 $\cald_R=2\pi T D$ of the baryonic membrane theory plasma
 for different quantizations
 of the gravitational dual
 (pseudo)scalars $\{\ln[v_1v_2^{-1}],b_1-b_2\}$: $\{\calo_2,\delta\calo_2^b\}$ (black,solid),
  $\{\calo_2,\delta\calo^b_1\}$ (black,dashed),
  $\{\calo_1,\delta\calo_2^b\}$ (blue,solid),
  $\{\calo_1,\delta\calo^b_1\}$ (blue,dashed).
At $q=0$, $\cald_R=\frac 32$ \eqref{an3}, while it vanishes in
the quantum critical regime $q\to q_{crit}$, correspondingly $T\to 0$.
}\label{figure3}
\end{figure}

For $q\in(0,q_{crit})$ \eqref{sol2} the $R$-charge
diffusion coefficient of the baryonic membrane theory plasma
is computed numerically, see fig.~\ref{figure3}.
We use the same color scheme as for the background in
figs.~\ref{figure1} and \ref{figure2}: the black curves correspond to
quantization of the background mode $\ln[v_1v_2^{-1}]\Leftrightarrow\calo_2$
and the blue curves to the identification
$\ln[v_1v_2^{-1}]\Leftrightarrow\calo_1$. Furthermore,
the (solid,dashed) curves represent the pseudoscalar
quantization $(b_1-b_2)\Leftrightarrow (\delta\calo_2^b,\delta\calo^b_1)$
correspondingly. In all cases we find the diffusion coefficient
$\cald>0$ for $q\in(0,q_{crit})$: the $R$-charge transport is stable. 
In the quantum critical regime $(q_{crit}-q)\ll q_{crit}$, equivalently
$T/\alpha\to 0 $ \eqref{tbaryonic},
$\cald\ \propto\ (q_{crit}-q)\ \propto T/\alpha$ (the solid curves and
the blue dashed curve),  except for the
quantization $\{\ln[v_1 v_2^{-2}],(b_1-b_2)\}\Leftrightarrow \{\calo_2,
\delta\calo^b_{1}\}$ (the dashed black curve) when 
$\cald\ \propto\ (q_{crit}-q)^2\ \propto (T/\alpha)^2$.

\subsection{Threshold instabilities from condensation of \texorpdfstring{$(b_1-b_2)$}{b1 - b2}}\label{axionkpt}

Consider spatially homogeneous and isotropic fluctuations of the bulk
pseudoscalars $b_1$ and $b_2$ about baryonic black membrane of section \ref{backpt}.
The corresponding equations of motion can be obtained from
\eqref{fl8}-\eqref{fl7} in the limit
\begin{equation}
\{w,k\}\to 0\,,
\eqlabel{homlimb}
\end{equation}
provided\footnote{We explicitly verified this.}
we replace \eqref{fl9} with $\cala^H_r=0$. The decoupled set of
linearized equations containing 
$\calb_1$ and $\calb_2$ is:
\begin{equation}
\begin{split}
&0=\calb_1''+\biggl(
\frac{v_2 g^4 s^2}{2r f}+\frac{r^3 v_2 s^2 q^2}{4f v_1^2}
+\frac{v_1^2 g^4 s^2}{4r f v_2}+\frac{r^3 s^2 q^2}{8f v_2}
-\frac{4 g^2 s^2}{r f v_1}-\frac{2 g^2 s^2}{r f v_2}
+\frac{9g^4 s^2}{4r f v_2 v_1^2}
-\frac{2 v_1'}{v_1}\\&+\frac 1r\biggr) \calb_1'
-\frac{s q r^2 (2 v_2^2-v_1^2)}{2f (2 v_2^2+v_1^2)} (\cala^H_t)'
-\frac{s r^2 q (2 v_2^2 v_1^2+v_1^4+6 v_2^2-3 v_1^2)}{4f (2 v_2^2+v_1^2)}
a\\&+ \biggl(
-\frac{(2 v_2^2+v_1^2+3) s^2g^4}{f v_2 r^2} 
+\frac{s^2 r^2 q^2 (2 v_2^2-v_1^2)^2}{2(v_1^2 f (2 v_2^2+v_1^2) v_2}
\biggr)\calb_1
+ \biggl(-\frac{(v_1^2+3) s^2g^4}{f v_2 r^2}
\\&-\frac{r^2 q^2 s^2 (2 v_2^2-v_1^2)}{2v_2 f (2 v_2^2+v_1^2)}\biggr) \calb_2\,,
\end{split}
\eqlabel{bsu1}
\end{equation}
\begin{equation}
\begin{split}
&0=\calb_2''+\biggl(
\frac{v_2 g^4 s^2}{2r f}+\frac{r^3 v_2 s^2 q^2}{4f v_1^2}
+\frac{v_1^2 g^4 s^2}{4r f v_2}+\frac{r^3 s^2 q^2}{8f v_2}
-\frac{4 g^2 s^2}{r f v_1}-\frac{2 g^2 s^2}{r f v_2}
+\frac{9g^4 s^2}{4r f v_2 v_1^2}
-\frac{2 v_2'}{v_2}\\&+\frac1r\biggr) \calb_2'
+\frac{q s r^2 v_2^2}{f (2 v_2^2+v_1^2)} (\cala^H_t)'
+\frac{q s r^2 v_2^2 (2 v_2^2+v_1^2+3)}{2f (2 v_2^2+v_1^2)} a
+\biggl(-\frac{2 (v_1^2+3) s^2 v_2g^4}{f v_1^2 r^2}
\\&-\frac{v_2 s^2 r^2 q^2 (2 v_2^2-v_1^2)}{v_1^2 f (2 v_2^2+v_1^2)}\biggr)\calb_1
- \frac{v_2 s^2 (4 v_2^2 g^4+2 g^4 v_1^2-r^4 q^2)}{f r^2 (2 v_2^2+v_1^2)}\calb_2\,,
\end{split}
\eqlabel{bsu2}
\end{equation}
\begin{equation}
\begin{split}
&0=a'+\biggl(
\frac{r (v_2')^2}{4v_2^2}+\frac{r (v_1')^2}{2v_1^2}
+\frac{r (g')^2}{g^2}+\frac{v_2'}{v_2}
+\frac{2v_1'}{v_1}\biggr) a
+\frac{6 s^2 g^4}{v_2 v_1^2 r^2 f} \cala^H_t
-\frac{2 s q}{v_2 v_1^2} \biggl(\calb_1'-\calb_2'\biggr)\,,
\end{split}
\eqlabel{bsu3}
\end{equation}
\begin{equation}
\begin{split}
&0=(\cala^H_t)''+\biggl(
\frac{r(g')^2}{g^2}+\frac{r (v_1')^2}{2v_1^2}
+\frac{r(v_2')^2}{4v_2^2}
-\frac{(2 v_2^2-v_1^2)v_2'}{v_2 (2 v_2^2+v_1^2)}
-\frac{2 v_1' v_1}{2 v_2^2+v_1^2}
\biggr) (\cala^H_t)'
\\&-\frac{(2 v_2^2 v_1^2+v_1^4+9) g^4 s^2}{f v_2 v_1^2 r^2} \cala^H_t
-\frac{(2 v_2^2-v_1^2-3) s q}{v_1^2 v_2} \calb_1'
+\frac{(v_1^2-3) s q}{v_1^2 v_2} \calb_2'
\\&-\frac{6 (v_2' v_2 v_1+v_2^2 v_1'+v_1' v_1^2)}
{(2 v_2^2+v_1^2) v_1} a+ \biggl(
\frac{2 s q (4 v_2^4+4 v_2^2 v_1^2-v_1^4)v_1'}{v_2 (2 v_2^2+v_1^2) v_1^3} 
-\frac{8 v_2' s q}{2 v_2^2+v_1^2}\biggr)\calb_1
\\&-\frac{2  s q (2 v_2' v_2+v_1' v_1)}{v_2 (2 v_2^2+v_1^2)} \calb_2\,,
\end{split}
\eqlabel{bsu4}
\end{equation}
where we introduced $a\equiv (\cala^0_t)'$.
\nxt In the UV, \ie as $r\to 0_+$, and with the identification\footnote{Likewise, we develop the UV expansions for the alternative quantization of the
background scalars $\ln[v_1 v_2^{-1}]\Longleftrightarrow\calo_1$.}
$\ln[v_1 v_2^{-1}]\Longleftrightarrow\calo_2$, 
\begin{equation}
\begin{split}
&\calb_1=b_{1,1} r+b_{1,2} r^2-\frac15 v_{1,2} b_{1,1} r^3
+\biggl(-\frac16 f_3 b_{1,1}-v_{1,2} b_{1,2}\biggr) r^4
+\biggl(
b_{1,5}+\biggl(-\frac{1}{70} q^2 b_{1,1}\\&
+\frac{12}{35} v_{1,2}^2
b_{1,1}\biggr) \ln r\biggr) r^5+\calo(r^6\ln r)\,,
\end{split}
\eqlabel{bb1un}
\end{equation}
\begin{equation}
\begin{split}
&\calb_2=-2 b_{1,1} r-2 b_{1,2} r^2+\frac{14}{5} v_{1,2} b_{1,1} r^3
+\biggl(\frac13 f_3 b_{1,1}+2 v_{1,2} b_{1,2}\biggr) r^4
+\biggl(-\frac{39}{20} v_{1,2}^2 b_{1,1}\\&+f_3 b_{1,2}-\frac15 q^2 b_{1,1}
+b_{1,5}-\frac32 b_{1,1} v_{1,4}+\biggl(
\frac{1}{35} q^2 b_{1,1}-\frac{24}{35} v_{1,2}^2 b_{1,1}\biggr) \ln r\biggr) r^5
+\calo(r^6\ln r)\,,
\end{split}
\eqlabel{bb2un}
\end{equation}
\begin{equation}
\begin{split}
&a=\frac{12}{5} q b_{1,1} r-(6 b_{1,1} q v_{1,2}+2 a^h_{t,4}) r^3
+\biggl(\frac12 f_3 q b_{1,1}-6 q v_{1,2} b_{1,2}\biggr) r^4+
+\calo(r^5\ln r)\,,\\
&\cala^H_t=
\frac35 q b_{1,1} r^2+2 q b_{1,2} r^3+a^h_{t,4} r^4+\calo(r^5)\,.
\end{split}
\eqlabel{baaht}
\end{equation}
Notice that $\lim_{r\to 0}a=0$ --- this ensures that the fluctuations
$\{\calb_i\,,\, \cala^0_t\,,\, \cala^H_t\}$ have the vanishing $R$-charge.  
In the quantization where $(b_1-b_2)$ is identified with the boundary
gauge theory operator $\delta\calo_2^b$ the coefficient $b_{1,1}$
is the source, while in the identification
$(b_1-b_2)\Longleftrightarrow\delta\calo^b_1$ the source term is $b_{1,2}$.
\nxt In the IR, \ie as $y\equiv 1-r\to 0$, 
\begin{equation}
\calb_1=b_{1;0}^h+\calo(y)\,,\quad \calb_2=b_{2;0}^h+\calo(y)\,,\quad
\cala^H_{t}=a^{h,h}_{t,1}\ y +\calo(y^2)\,,\quad a=a_{0}^h+\calo(y)\,.
\eqlabel{b1b2hb}
\end{equation}

\begin{figure}[t]
\begin{center}
\psfrag{on}[c]{{$\delta\calo_2^b\propto b_{1,2} $}}
\psfrag{oa}[t]{{$\delta\calo^b_1\propto b_{1,1}$}}
\psfrag{q}{{$q/q_{crit}$}}
  \includegraphics[width=3in]{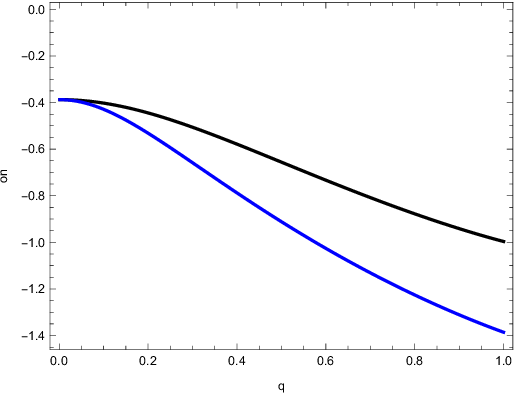}\
  \includegraphics[width=3in]{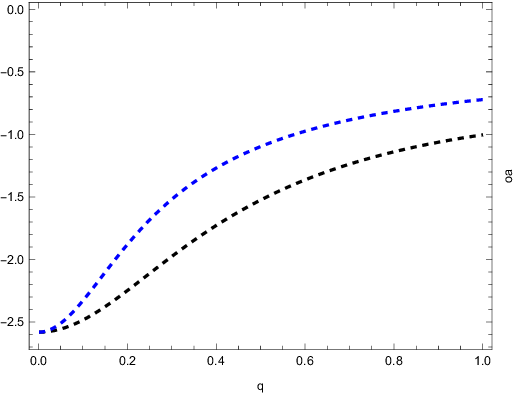}  
\end{center}
 \caption{We test the instability of the baryonic black
 membranes due to the condensation of the
 neutral $(b_1-b_2)$ mode: the instability
 would be signaled by the divergence of the
 response of the corresponding operator for
 its fixed source, as we vary $q/q_{crit}$.
 The color coding is as in
 fig.~\ref{figure3}.
}\label{figure3a}
\end{figure}

Following \cite{Buchel:2019pjb}, to identify the onset of the instability
associated with the condensation of $\delta\calo_2^b$ ( or $\delta\calo^b_1$ )
we keep fixed the source term of the operator, $b_{1,1}=1$ (or $b_{1,2}=1$),
and scan $q$ (correspondingly ${T}$, see \eqref{tbaryonic})
looking for the divergence of the expectation value of the corresponding
operator $\langle\delta\calo_2^b\rangle\propto b_{1,2}$ ( or $\langle\delta\calo^b_1\rangle\propto b_{1,1}$ ).
A divergence signals the presence of a homogeneous and isotropic
normalizable mode of the fluctuations of $(b_1-b_2)$ --- the threshold
for the instability.
Results of such scans are presented in fig.~\ref{figure3a}:
there are no divergences of the expectation values of $\delta \calo^b$ operators.

\section{Reissner–Nordstr\"om black membrane}\label{rnbm}

Besides the baryonic black membranes described in the previous section, the truncation described in section \ref{effact} also features black membranes of the Reissner--Nordström type charged under the “non-topological” R-charge $U(1)$. In this section we describe these backgrounds and analyze whether they are stable with respect to \textit{baryonic} charge density fluctuations, the threshold instability, and the superconducting instability.

\subsection{Background}\label{rnbackground}

Minimal $\caln=4$ $D=4$ gauged supergravity can be obtained
from \eqref{actionckv} truncating the complex scalars $t^1=t^2=1+i 0$,
and the massive 2-form $B$ (equivalently the massive vector $A^H$ in \eqref{ahaction}).
This is achieved with the gauge fields $\calf^0\,, \calf^{1}\,, \calf^2$
satisfying (see also \cite{Gauntlett:2009zw})
\begin{equation}
\calf^1=\calf^2=\star \calf^0\,.
\eqlabel{f0f1}
\end{equation}
We will be interested in $U(1)_R$-charged black membrane solutions of the
resulting effective action. Using the metric ansatz as in \eqref{4dm},
the background geometry is given by
\begin{equation}
\begin{split}
&ds_4^2=-\frac{4\alpha^2 f}{r^2}\ dt^2+\frac{4\alpha^2}{r^2} \left(dx_1^2+dx_2^2\right)
+ \frac{1}{4r^2 f}\ dr^2\,,\qquad f=1-r^3 (1+q^2)+q^2 r^4\,,\\
&\calf^0=a_0'\ dr\wedge dt\,,\qquad \calf^1=\calf^2=\calb\ dx_1\wedge dx_2\,,\qquad
a_0=2q \alpha(1-r)\,,\\
&\calb=-8\alpha^2 q\,,\qquad  b_1=b_2=0\,,\qquad v_1=v_2=1\,.
\end{split}
\eqlabel{rn}
\end{equation}
From \eqref{rn} we extract an $R$-charge chemical potential of $\mu_R=2q\alpha$ and
a Hawking temperature of the black membrane $T=\alpha(3-q^2)/\pi$, leading to
\begin{equation}
\frac{T}{\mu_R}=\frac{3-q^2}{2\pi q}\,.
\eqlabel{murt}
\end{equation}
Here the extremal $AdS_2\times \reals^2$ limit is reached as
$q\to q_{crit}=\sqrt{3}$.

\subsection{Fluctuations}\label{rnfluc}

In this section we discuss fluctuations of the baryonic charge density
fluctuations about the RN black membrane of section \ref{rnbackground}.
Within the effective action \eqref{actionckv} and \eqref{ahaction}
we consider linearized
fluctuations
\begin{equation}
\begin{split}
&A^0=\delta A^0_{x_1}\ d x_1+\delta A^0_r\ dr\,,\qquad
A_H= \delta A^H_{x_1}\ d x_1\\
&\delta A^i = \delta A^i_t\ dt+\delta A^i_{x_2}\ dx_2+\delta A^i_r\ dr\,,\qquad b_i=\delta b_i\,,
\end{split}
\eqlabel{fldecrn}
\end{equation}
where
\begin{equation}
\delta A^{i}_{t,x_2,r}=e^{-i w t +i k x_2}  \cala^{i}_{t,2,r}(r) \,, \quad
\delta A^{0,H}_{x_1,r}=e^{-i w t +i k x_2} \cala^{0,H}_{1,r}(r) \,, \quad
\delta b_i= e^{-i w t +i k x_2} \calb_i(r) \,,
\end{equation}
about the black membrane background \eqref{rn}. It is straightforward
to verify that the set \eqref{fldecrn} will decouple from the remaining
fluctuations in the helicity-0 (the sound channel) sector.
We use the bulk gauge transformations \eqref{gauge} to set
\begin{equation}
\delta \cala^i_r=\delta \cala^0_r\equiv 0\,.
\eqlabel{fixgaugern}
\end{equation}
The equations of motion for the remaining fluctuations
take the form
\begin{equation}
\begin{split}
0=&\biggl(\cala^1_t-\cala^2_t\biggr)''
+\biggl(
-\frac{c_3'}{c_3}-\frac{c_1'}{c_1}+\frac{2c_2'}{c_2}\biggr) \biggl(\cala^1_t
-\cala^2_t\biggr)'-\frac{c_3^2 k^2}{c_2^2} \biggl(\cala^1_t-\cala^2_t\biggr)
-\frac{c_3^2 k w}{c_2^2} \biggl(\cala^1_2\\
&-\cala^2_2\biggr)-\frac12 a_0' (4 \calb_1'-\calb_2')\,,
\end{split}
\eqlabel{eomrn1}
\end{equation}
\begin{equation}
\begin{split}
0=&\biggl(\cala^1_2-\cala^2_2\biggr)''+\biggl(
-\frac{c_3'}{c_3}+\frac{c_1'}{c_1}\biggr) \biggl(
\cala^1_2-\cala^2_2\biggr)'
+\frac{c_3^2 w^2}{c_1^2} \biggl(\cala^1_2-\cala^2_2\biggr)
+\frac{c_3^2 k w}{c_1^2} \biggl(\cala^1_t-\cala^2_t\biggr)\,,
\end{split}
\eqlabel{eomrn2}
\end{equation}
\begin{equation}
\begin{split}
0=&\biggl(\cala^H_1+2 \cala^0_1\biggr)''
+\biggl(-\frac{c_3'}{c_3}+\frac{c_1'}{c_1}\biggr)
\biggl(\cala^H_1+2 \cala^0_1\biggr)'+\frac{c_3^2 (c_2^2 w^2-c_1^2 k^2)}{c_1^2 c_2^2}
\biggl(\cala^H_1+2 \cala^0_1\biggr)\,,
\end{split}
\eqlabel{eomrn3}
\end{equation}
\begin{equation}
\begin{split}
0=&\biggl(\cala^H_1+\frac32 \cala^0_1\biggr)''
+\biggl(-\frac{c_3'}{c_3}+\frac{c_1'}{c_1}\biggr)
\biggl(\cala^H_1+\frac32 \cala^0_1\biggr)'
+\biggl(\frac{c_3^2 w^2}{c_1^2}-\frac{c_3^2 k^2}{c_2^2}\biggr) \biggl(
\cala^H_1+\frac32 \cala^0_1\biggr)
\\&-12 c_3^2 \cala^H_1-\frac{ic_3 a_0' k}{2c_1}  (2 \calb_1+\calb_2)\,,
\end{split}
\eqlabel{eomrn4}
\end{equation}
\begin{equation}
\begin{split}
0=&\calb_1''+\biggl(
-\frac{c_3'}{c_3}+\frac{c_1'}{c_1}+\frac{2c_2'}{c_2}\biggr) \calb_1'
+\biggl(-24 c_3^2+\frac{c_3^2 w^2}{c_1^2}
-\frac{c_3^2 k^2}{c_2^2}-\frac{2(a_0')^2}{3c_1^2}\biggr) \calb_1
-\frac{2a_0'}{3c_1^2} \biggl(\cala^1_t\\
&-\cala^2_t\biggr)'
+\biggl(-16 c_3^2-\frac{4(a_0')^2}{3c_1^2}\biggr) \calb_2+\frac{2ic_3 a_0' k}{3c_1 c_2^2}  \cala^H_1 \,,
\end{split}
\eqlabel{eomrn5}
\end{equation}
\begin{equation}
\begin{split}
0=&\calb_2''+\biggl(
-\frac{c_3'}{c_3}+\frac{c_1'}{c_1}+\frac{2 c_2'}{c_2}\biggr)
\calb_2'+\biggl(-8 c_3^2+\frac{c_3^2 w^2}{c_1^2}-\frac{c_3^2 k^2}{c_2^2}
+\frac{2(a_0')^2}{3c_1^2}\biggr) \calb_2+\frac{a_0'}{3c_1^2}
\biggl(\cala^1_t\\&-\cala^2_t\biggr)'
+\biggl(-32 c_3^2-\frac{8(a_0')^2}{3c_1^2}\biggr) \calb_1+\frac{2ic_3a_0'k}{3c_1 c_2^2} \cala^H_1 \,,
\end{split}
\eqlabel{eomrn6}
\end{equation}
along with the constraint
\begin{equation}
\begin{split}
0=&\biggl(\cala^1_t-\cala^2_t\biggr)'+\frac{c_1^2 k}{c_2^2 w}
\biggl(\cala^1_2-\cala^2_2\biggr)'-\frac12 a_0' (4 \calb_1-\calb_2)\,,
\end{split}
\eqlabel{eomrn7}
\end{equation}
where, compare with \eqref{rn},
\begin{equation}
c_1=\frac{2\alpha \sqrt f}{r}\,,\qquad
c_2=\frac{2\alpha}{r}\,,\qquad c_3=\frac{1}{2r\sqrt f }\,.
\eqlabel{defcfrn}
\end{equation}
We explicitly verified that \eqref{eomrn7} is consistent with
\eqref{eomrn1}-\eqref{eomrn6}.
Fluctuations of the $U(1)$ baryonic current dual gauge potential $A^1-A^2$,
see \cite{Cassani:2012pj},  excite
the axions $b_1$ and $b_2$ (see \eqref{eomrn5} and \eqref{eomrn6}).
The holographic spectroscopy relates
these scalars  to the boundary gauge theory operators
$\delta\calo_\Delta^b$ of conformal dimension $\Delta$ as in
table \ref{table2}.

Notice that \eqref{eomrn3} can be solved trivially with
\begin{equation}
\cala_1^H=-2\cala^0_1\,.
\eqlabel{solveaH1}
\end{equation}
Similar to  section \ref{fluckpt}, we introduce
\begin{equation}
Z\equiv \kk\ \biggl(\cala^1_t-\cala^2_t\biggr)
+ \ww\ \biggl(\cala^1_2-\cala^2_2\biggr)\,.
\eqlabel{zrn}
\end{equation}
We use the constraint \eqref{eomrn7}  to obtain from \eqref{eomrn1}-\eqref{eomrn6} a decoupled
set of the second-order equations for
\begin{equation}
\{\ Z\,,\ \cala^0_1\,,\  \calb_1\,,\ \calb_2\ \}\,.
\eqlabel{setrn}
\end{equation}
Solutions of the resulting equations with appropriate boundary conditions
determine the spectrum of $R$-charged black membranes quasinormal 
modes --- equivalently
the physical spectrum of linearized fluctuations in membrane
gauge theory plasma with
a baryonic chemical potential. Following \cite{Kovtun:2005ev,Son:2002sd}
we impose the incoming-wave boundary conditions at the black membrane
horizon, and 'normalizability'
at asymptotic $AdS_4$ boundary.
Focusing on the $\Re[\ww]=0$  diffusive branch, and
introducing
\begin{equation}
\begin{split}
&Z=\left(1-r\right)^{-i\ww/2}\ z\,,\ \ \cala^0_1=i (1-r)^{-i\ww/2}\ a\,,\ \ \calb_i=(1-r)^{-i\ww/2}\ B_i\,,\  \ 
\ww=-i v\ \kk\,,
\end{split}
\eqlabel{incomern}
\end{equation}
we solve the fluctuation equations  subject to the asymptotics: 
\nxt in the UV, \ie as $r\to 0_+$, and with the
identification\footnote{Likewise, we develop the UV expansions for the
alternative quantization of the fluctuation $(b_1-b_2)$:
$\{\delta\calo^b_1\}$.} $(b_1-b_2)\Longleftrightarrow\delta\calo_2^b$, 
\begin{equation}
\begin{split}
&z=\kk r -\frac 12 \kk^2 v r^2+\calo(r^3)\,,\quad a=a_4 r^4+\calo(r^5)\,,
\quad B_2=-2 b_{1,2} r^2+b_{1,2} \kk r^3 v+\calo(r^4)\,,\\
&B_1=b_{1,2} r^2-\frac12 b_{1,2} \kk v r^3+
\frac{b_{1,2}\kk}{24}\biggl(\kk (v^2+1) (q^2+1)^2-8 \kk (v^2+1) (q^2+1)+19 \kk v^2+16 \kk\\
&-6 v\biggr)r^4+b_{1,5} r^5+\calo(r^6)\,,
\end{split}
\eqlabel{uv1rn}
\end{equation}
specified, for a fixed background and a momentum $\kk$,  by
\begin{equation}
\biggl\{\
v\,,\ a_{4}\,,\ b_{1,2}\,,\ b_{1,5}
\
\biggr\}\,;
\eqlabel{fluvrn}
\end{equation}
\nxt in the IR, \ie as $y\equiv 1-r\to 0_+$,
\begin{equation}
\begin{split}
&z=z^h_0+\calo(y)\,,\qquad a=a^h_0+\calo(y)\,,\qquad
B_i=b^h_{i;0}+\calo(y)\,,
\end{split}
\eqlabel{ir1rn}
\end{equation}
specified  by
\begin{equation}
\biggl\{\
z^h_0\,,\ a^{h}_{0}\,,\
b^{h}_{1;0}\,,\ b^{h}_{2;0} 
\
\biggr\}\,.
\eqlabel{flirrn}
\end{equation}
Note that in total we have $4+4=8$ parameters, see \eqref{fluvrn} and
\eqref{flirrn},
which is precisely what is necessary to identify a solution of a coupled system of
4 second-order ODEs for $\{z, a, B_1, B_2\}$. Furthermore, without
the loss of generality we normalized the solutions as in \eqref{normaliz}.

Once we fix the background, and solve the fluctuation equations of motion,
we obtain $v=v(\kk)$. Given $v$ we extract the baryonic charge
diffusion coefficient $\cald$ as in \eqref{defd}.
For general values of $q$ we have to solve the fluctuation equations
numerically. In the limit $q=0$ the diffusion coefficient can
be computed analytically, see \eqref{an3}.

\begin{figure}[t]
\begin{center}
\psfrag{d}{{$\cald_B$}}
\psfrag{q}{{$q/q_{crit}$}}
  \includegraphics[width=4in]{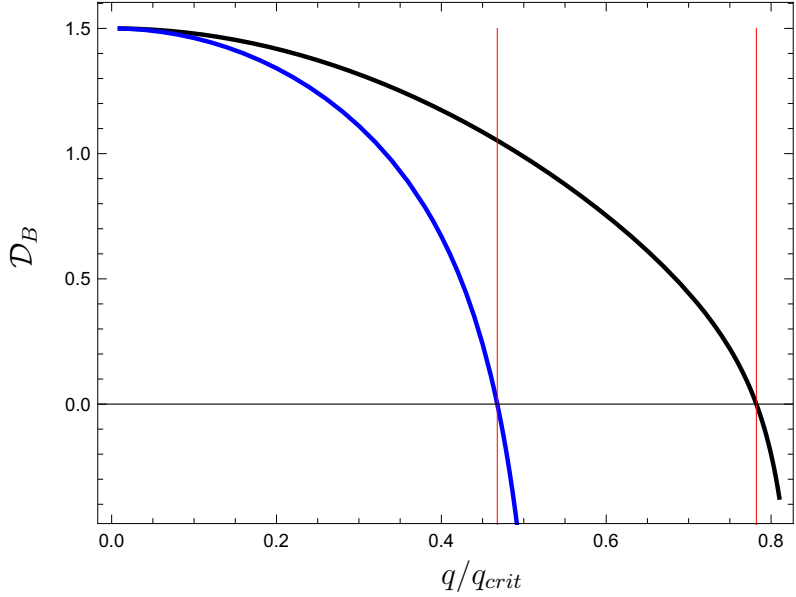}
\end{center}
 \caption{Dimensionless baryonic charge diffusion
 coefficient $\cald_B=2\pi T D$ of the
 $R$-charged membrane theory plasma for different quantizations
 of the gravitational dual pseudoscalar
 $(b_1-b_2)$: $\{\delta\calo_2^b\}$ (black),
  $\{\delta\calo^b_1\}$ (blue). The vertical red lines indicate
  the onset of the baryonic charge
  clumping instability, see \eqref{tcritb}.
}\label{figure4}
\end{figure}

Results for the baryonic charge diffusion coefficient
of the  $R$-charged membrane theory plasma  are presented
in fig.~\ref{figure4}. The black curve corresponds to the axion
$(b_1-b_2)$ identification with $\delta\calo^b_2$ boundary
operator, and the blue curve corresponds to the quantization
$(b_1-b_2)\Leftrightarrow \delta\calo^b_1$. 
Note that in both cases $\cald$ vanishes at certain value of $q/q_{crit}$
(represented by vertical red lines),
correspondingly the temperature, see \eqref{murt},
\begin{equation}
\frac{T}{\mu_R}\bigg|_{(b_1-b_2)\Leftrightarrow\delta\calo_2^b}^{black}=0.13(7)\,,\qquad
\frac{T}{\mu_R}\bigg|_{(b_1-b_2)\Leftrightarrow\delta\calo^b_1}^{blue}=0.46(0)\,,
\eqlabel{tcritb}
\end{equation}
and becomes negative at yet lower temperatures. The negativity
of the diffusion coefficient indicates unstable transport, physically
realized as a baryonic charge clumping.

\subsection{Threshold instabilities from condensation of \texorpdfstring{$(b_1-b_2)$}{b1 - b2}}\label{axionrn}

Consider spatially homogeneous and isotropic fluctuations of the bulk
pseudoscalars $b_1$ and $b_2$ about $R$-charged black membrane \eqref{rn}.
The corresponding equations of motion can be obtained from
\eqref{eomrn1}-\eqref{eomrn6} in the limit
\begin{equation}
\{w,k\}\to 0\,,
\eqlabel{homlim}
\end{equation}
provided
we drop\footnote{Much like
in the related analysis in \cite{Buchel:2025jup}, this constraint
equation is multiplied by $w$, and is trivially satisfied
for spatially homogeneous and isotropic fluctuations.}
the constraint \eqref{eomrn7}. 
Solving \eqref{eomrn1} in the limit \eqref{homlim} we find
\begin{equation}
\biggl(\cala^1_t-\cala^2_t\biggr)'={\rm const}+q(\calb_2-4\calb_1)\,.
\eqlabel{atsol}
\end{equation}
No matter what quantization is used for $(b_1-b_2)$ bulk pseudoscalar,
the constant in \eqref{atsol} is related to the baryonic charge
of the black membrane (in addition to the $R$-charge determined by $q$).
Thus, we must set ${\rm const}=0$ in \eqref{atsol}.
Using \eqref{atsol}, we identify the decoupled set of linearized equations for
$\calb_1$ and $\calb_2$ from \eqref{eomrn5} and \eqref{eomrn6} in the limit
\eqref{homlim}:
\begin{equation}
\begin{split}
&0=\calb_1''+\biggl(
\frac{f'}{f}-\frac2r\biggr) \calb_1'-\frac{2 q^2 r^4+6}{r^2 f} \calb_1
-\frac{q^2 r^4+4}{r^2 f} \calb_2 \,,\\
&0=\calb_2''+\biggl(
\frac{f'}{f}-\frac2r\biggr) \calb_2'-\frac{4-q^2 r^4}{2r^2 f} \calb_2
- \frac{2(q^2 r^4+4)}{r^2 f}\calb_1 \,,
\end{split}
\eqlabel{brn}
\end{equation}
where from \eqref{rn} $f=1-r^3(1+q^2)+q^2 r^4$.
\nxt In the UV, \ie as $r\to 0_+$, the general solution of \eqref{brn}
takes the form
\begin{equation}
\begin{split}
&\calb_1=b_{1,1}\ r+b_{1,2}\ r^2+
\frac16 (q^2+1) b_{1,1}\ r^4
+\biggl(b_{1,5}+\frac17  q^2 b_{1,1}\ \ln r\biggr)\ r^5+\calo(r^6)\,,
\end{split}
\eqlabel{b1uvrn}
\end{equation}
\begin{equation}
\begin{split}
&\calb_2=-2 b_{1,1}\ r-2 b_{1,2}\ r^2-
\frac13 (q^2+1) b_{1,1}\ r^4+\biggl(
\frac34 q^2 b_{1,1}-(q^2+1) b_{1,2}+b_{1,5}\\
&+\frac17  q^2 b_{1,1}\ \ln r\biggr)
\ r^5+\calo(r^6)\,.
\end{split}
\eqlabel{b2uvrn}
\end{equation}
In the quantization where $(b_1-b_2)$ is identified with the boundary
gauge theory operator $\delta\calo_2^b$ the coefficient $b_{1,1}$
is the source, while in the identification
$(b_1-b_2)\Longleftrightarrow\delta\calo^b_1$ the source term is $b_{1,2}$.
\nxt In the IR, \ie as $y\equiv 1-r\to 0$, 
\begin{equation}
\calb_1=b_{1;0}^h+\calo(y)\,,\qquad \calb_2=b_{2;0}^h+\calo(y)\,.
\eqlabel{b1b2hrn}
\end{equation}

\begin{figure}[t]
\begin{center}
\psfrag{on}[c]{{$\langle\delta\calo_2^b\rangle^{-1} $}}
\psfrag{oa}[t]{{$\langle\delta\calo^b_1\rangle^{-1}$}}
\psfrag{q}{{$q/q_{crit}$}}
  \includegraphics[width=3in]{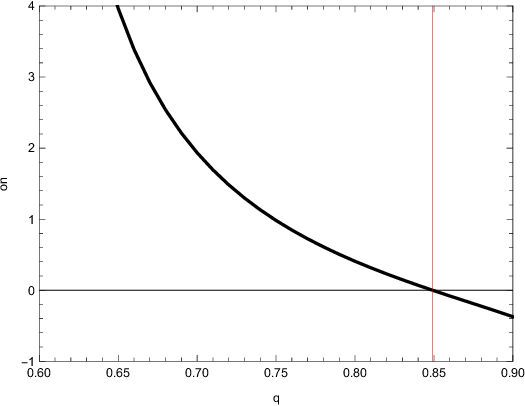}\
  \includegraphics[width=3in]{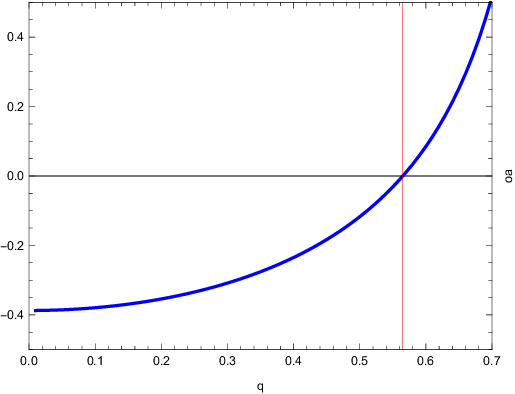}  
\end{center}
 \caption{Divergence of the expectation value of the
 operator dual to spatially
 homogeneous and isotropic fluctuations of the bulk
 pseudoscalar $(b_1-b_2)$
 for different quantizations: 
 $\{\delta\calo_2^b\}$ (black, the left panel), and 
  $\{\delta\calo^b_1\}$ (blue, the right panel). The vertical red lines indicate
  the onset of the instability, see \eqref{tcritbsus}.
}\label{figure5}
\end{figure}

Following \cite{Buchel:2019pjb}, to identify the onset of instability
associated with the condensation of $\delta\calo_2^b$ ( or $\delta\calo^b_1$ )
we keep fixed the source term of the operator, $b_{1,1}=1$ (or $b_{1,2}=1$),
and scan $q$ (correspondingly ${T}/{\mu_R}$, see \eqref{murt})
looking for the divergence of the expectation value of the corresponding
operator $\langle\delta\calo_2^b\rangle\propto b_{1,2}$ ( or $\langle\delta\calo^b_1\rangle\propto b_{1,1}$ ).
A divergence signals the presence of a homogeneous and isotropic
normalizable mode of the fluctuations of $(b_1-b_2)$ --- the threshold
for the instability.
Results of such scans are presented in fig.~\ref{figure5}.
We observe the onset of the instabilities at temperatures 
\begin{equation}
\frac{T}{\mu_R}\bigg|_{(b_1-b_2)\Leftrightarrow\delta\calo_2^b}^{black}=0.09(1)\,,\qquad
\frac{T}{\mu_R}\bigg|_{(b_1-b_2)\Leftrightarrow\delta\calo^b_1}^{blue}=0.33(2)\,,
\eqlabel{tcritbsus}
\end{equation}
represented by the vertical red lines for the corresponding values of
$q/q_{crit}$. The temperatures \eqref{tcritbsus} are lower
for the corresponding quantizations of $(b_1-b_2)$ gravitational
pseudoscalar then \eqref{tcritb} --- thus, the baryonic charge clumping
occurs prior to the condensation of $\delta\calo^b$ in the RN black
membrane background. New phases of the $R$-charged black membranes in
our model with $\langle\delta\calo^b\rangle\ne 0$ will be discussed
elsewhere.

\subsection{Superconducting instability}\label{rnsuper}

In this section we complete discussion of the potential instabilities of the
$R$-charged black membranes. The effective action \eqref{actionckv}
reviewed in section \ref{effact} does not contain any $U(1)_R$
charged matter. The most general $\caln=2$ gauged supergravity
obtained from the consistent truncation of M-theory on $M^{1,1,0}$
coset includes a pair of $R$-charged real scalars $\xi^0$ and $\tilde\xi^0$
\cite{Cassani:2012pj}. It is technically more transparent to discuss
this charged sector using the effective action of \cite{Gauntlett:2009zw}.

From \cite{Gauntlett:2009zw}, the effective action is
\begin{equation}
\begin{split}
S=\int d^4 &x \sqrt{-g} \biggl[
R-24(\nabla U)^2-\frac 32 (\nabla V)^2-6 \nabla U\cdot \nabla V\\
&-\frac 32 e^{-4 U-2V}(\nabla h)^2-\frac 32 e^{-6U}|D\chi|^2
-\frac14 e^{6U+3V} F_{\mu\nu}F^{\mu\nu} \\
&-\frac{1}{12}e^{12U} H_{\mu\nu\rho}H^{\mu\nu\rho}-\frac 34
e^{2U+V}H_{\mu\nu}H^{\mu\nu}+48e^{-8U-V}-6 e^{-10U+V}\\
&-24h^2 e^{-14U-V}-18(1+h^2+|\chi|^2)^2 e^{-18U-3V}-24e^{-12U-3V}|\chi|^2
\biggr]\\
+\int \biggl[&-3h H_2\wedge H_2 +3h^2 H_2 \wedge F_2-h^3 F_2\wedge F_2
+6 A_1\wedge H_3\\
&-\frac {3i}{4} H_3\wedge \left(\chi^* D\chi-\chi (D\chi)^*\right)\biggr]\,,
\end{split}
\eqlabel{jg1}
\end{equation}
where
\begin{equation}
H_3=dB_2\,,\ H_2=dB_1+2B_2+hF_2\,,\ F_2=dA_1\,,\ D\chi\equiv d \chi
-4 i A_1\chi\,.
\eqlabel{jg2}
\end{equation}
Effective action \eqref{jg1} can be matched to the CKV effective action of
\cite{Cassani:2012pj} 
as follows:
\begin{equation}
\begin{split}
&U=-\frac 13 \phi\,,\qquad V=\frac 23\phi +\ln v\,,\qquad  {\rm where}\qquad v_1=v_2=v_3\equiv v\,,\\
&B_2=B\,,\qquad A_1=A^0\,,\qquad B_1=-A\,,\qquad {\rm where}\qquad A^1=A^2=A^3\equiv A\,,\\
&\chi=\frac{1}{\sqrt 3}\left(\xi^0+i \tilde\xi^0\right)\,,\qquad h=b\,,\qquad
{\rm where}\qquad b_1=b_2=b_3\equiv b\,.
\end{split}
\eqlabel{jg3}
\end{equation}

From \eqref{jg1}, quadratic effective action $S_\chi$ for the 
complex scalar $\chi\equiv \eta e^{i\Theta}$, dual to
the boundary membrane operator $\calo_\chi$ of conformal
dimension $\Delta=5$ and $R$-charge $R(\chi)=4$,
about $R$-charged black membrane
\eqref{rn} takes the form
\begin{equation}
\begin{split}
S_\chi=&\int dx^4 \sqrt{-g} \biggl[-\frac 32 |D\chi|^2 -60 |\chi^2| \biggr]\\
=&\int dx^4 \sqrt{-g} \biggl[-\frac 32 (\nabla\eta)^2
-\frac 32 \eta^2 \left(\nabla\Theta-4 A^0\right)^2-60 \eta^2 \biggr]\,.
\end{split}
\eqlabel{jg4}
\end{equation}
The phase $\Theta$ of the complex scalar $\chi$ can be gauged away,
and we arrive at the linearized equation for $\eta$
in the RN black membrane background \eqref{rn}:
\begin{equation}
0=\eta''+\left(\frac{f'}{f}-\frac2r\right)\ \eta'
-\frac{2(q (r^3-r^2)+5f)}{f^2r^2}\ \eta\,,
\eqlabel{jg5}
\end{equation}
where $f$ is specified in \eqref{rn}.
\nxt In the UV, \ie as $r\to 0_+$, the general solution of \eqref{jg5}
takes form
\begin{equation}
\begin{split}
\eta=&\eta_{-2}\ r^{-2}+\frac15 \eta_{-2} q-\frac16 \eta_{-2} (2 q^2+q+2)\ r
+\frac15 q^2 \eta_{-2}\ r^2\\
&+\frac{1}{150} \eta_{-2} q (20 q^2-11 q+20)\ r^3
+\frac{1}{180} \eta_{-2} (10 q^4-83 q^3+30 q^2-35 q+10)\ r^4
\\&+\biggl(\eta_5+\frac{4}{175} q^2 \eta_{-2} (5 q^2-7 q+5)\ \ln r\biggr)\
r^5+\calo(r^6)\,.
\end{split}
\eqlabel{jg6}
\end{equation}
In \eqref{jg6} the coefficients $\eta_{-2}$
is the source for the dual operator $\calo_\chi$, 
while its  expectation value $\langle\delta\calo_\chi\rangle\propto \eta_5$.
\nxt In the IR, \ie as $y\equiv 1-r\to 0$, 
\begin{equation}
\eta=\eta_{0}^h+\calo(y)\,.
\eqlabel{jg7}
\end{equation}

\begin{figure}[t]
\begin{center}
\psfrag{o}[c]{{$\eta_5$}}
\psfrag{q}{{$q/q_{crit}$}}
  \includegraphics[width=4in]{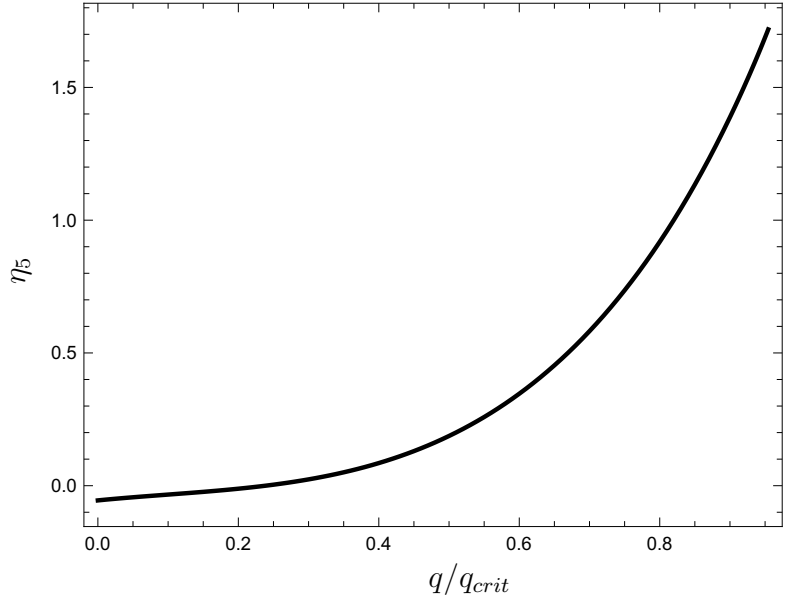}\
\end{center}
 \caption{The normalizable coefficient $\eta_5$ does not diverge at finite
 values of $q$, correspondingly $\frac{T}{\mu_R}\ne 0$ ---
 the dual operator $\calo_\chi$ does not condense at nonzero temperature.
}\label{figure6}
\end{figure}

Once again, to identify the onset of instability \cite{Buchel:2019pjb}
associated with the condensation of $\calo_\eta$
we keep fixed its source  term, $\eta_{-2}=1$,
and scan $q$ (correspondingly ${T}/{\mu_R}$, see \eqref{murt})
looking for the divergence of the expectation value coefficient $\eta_5$.
The result of such scan is presented in fig.~\ref{figure6}.
The lack of the instability is consistent with the analysis
of \cite{Denef:2009tp} (once we match the conventions).

\section*{Acknowledgments} We would like to 
thank Jerome  Gauntlett and Chiara Toldo for useful correspondence, and Igor Klebanov
for stimulating our interest in this project. 
AB gratefully acknowledges support from the Simons Center for Geometry and Physics,
Stony Brook University, at which some of the research for this paper was performed,
as well as the kind hospitality and stimulating research environments of
Centro de Ciencias de Benasque and the Institute for Nuclear Theory at the
University of Washington. 
This research was supported in part by the INT's U.S.
Department of Energy grant No. DE-FG02-00ER41132.
AB's work was further supported by NSERC through the Discovery Grants program.

\bibliographystyle{JHEP}
\bibliography{kpt}

\end{document}